\documentclass[twocolumn]{revtex4}

\usepackage{graphicx}
\usepackage{listings}
\usepackage{slashed}
\usepackage{MnSymbol}
\newcommand{\ac}[1]{#1}

\newcommand{\acs}[1]{#1}
\newcommand{\acp}[1]{#1}

\newcommand{\texorpdfstring}[2]{#1}
\newcommand{\burl}[1]{\url{#1}}

\newcommand{\be}{\begin{equation}}
\newcommand{\ee}{\end{equation}}
\newcommand{\ba}{\begin{eqnarray}}
\newcommand{\ea}{\end{eqnarray}}

\newcommand{\pderiv}[2]{\frac{\partial #1}{\partial #2}}

\newcommand{\bra}[1]{\langle #1|}
\newcommand{\ket}[1]{|#1\rangle}
\newcommand{\braket}[2]{\langle #1|#2\rangle}

\renewcommand{\vec}[1]{ {\bf #1}}

\def\ie{{\em i.e.}\ }

\newcommand{\Aphi}{A_\phi(\rho)}

\newcommand{\bhattext}[1]{${\bf \hat{#1}}$}

\newcommand{\mean}[1]{\langle#1\rangle}

\newcommand{\Paulispin}{\frac{1}{4} {\rm tr}e^{\frac{1}{2}\int_0^T d\tau \sigma_{\mu \nu}F^{\mu \nu}(x_{\rm CM}+x(\tau))}}

\def\GeV1{{\rm GeV}^{-1}}

\begin{document}
\title{Casimir Interactions between Magnetic Flux Tubes in a Dense Lattice} 
\author{Dan Mazur}
\email{daniel.mazur@mcgill.ca}
\affiliation{McGill High Performance Computing Centre, 
McGill University, 1100 Rue Notre-Dame Ouest, Montreal, QC H3C 1K3}
\author{Jeremy S. Heyl}
\email{heyl@phas.ubc.ca}
\affiliation{Department of Physics and Astronomy,
University of British Columbia, Vancouver BC V6T 1Z1 Canada; Canada Research Chair}
\begin{abstract}
  We use the worldline numerics technique to study a cylindrically 
  symmetric model of magnetic flux tubes in a dense lattice and the 
  non-local Casimir forces acting between regions of magnetic flux. 
  Within a
  superconductor the magnetic field is constrained within magnetic
  flux tubes and if the background magnetic field is on the order 
  the quantum critical field strength,
  $B_k = \frac{m^2}{e} = 4.4 \times 10^{13}$ Gauss,  
  the magnetic field is likely to vary
  rapidly on the scales where \acs{QED} effects are important.  In
  this paper, we construct a cylindrically symmetric toy model of a
  flux tube lattice in which the non-local influence of \acs{QED} on
  neighbouring flux tubes is taken into account.  We compute the
  effective action densities using the worldline numerics technique.
  The numerics predict a greater effective energy density in the
  region of the flux tube, but a smaller energy density in the regions
  between the flux tubes compared to a locally-constant-field
  approximation.  We also compute the interaction energy between a
  flux tube and its neighbours as the lattice spacing is reduced from
  infinity. Because our flux tubes exhibit compact support, this
  energy is entirely non-local and predicted to be zero in local
  approximations such as the derivative expansion. This Casimir-Polder
  energy can take positive or negative values depending 
  on the distance between the
  flux tubes, and it may cause the flux tubes in neutron stars
  to form bunches.  

  In addition to the above results we also discuss two important
  subtleties of determining the statistical uncertainties within the 
  worldline numerics technique.
  Firstly, the distributions generated
  by the worldline ensembles are highly non-Gaussian, and so the
  standard error in the mean is not a good measure of the statistical
  uncertainty.  Secondly, because the same ensemble of worldlines is
  used to compute the Wilson loops at different values of $T$ and
  $x_\mathrm{ cm}$, the uncertainties associated with each computed
  value of the integrand are strongly correlated. We recommend a form
  of jackknife analysis which deals with both of these problems.  
\end{abstract}

\maketitle

\section{Introduction}

In this contribution, we make use of a numerical
technique which can be used to compute the effective actions of
external field configurations.  The technique, called either worldline
numerics or the Loop-Cloud Method, was first used by Gies and
Langfeld~\cite{Gies:2001zp} and has since been applied to computation
of effective actions
\cite{Gies:2001tj,Langfeld:2002vy,Gies:2005sb,Gies:2005ym,Dunne:2009zz}
and Casimir energies
\cite{Moyaerts:2003ts,Gies:2003cv,PhysRevLett.96.220401}. More
recently, the technique has also been applied to pair
production~\cite{2005PhRvD..72f5001G} and the vacuum polarization
tensor~\cite{PhysRevD.84.065035}.  Worldline numerics 
is able to compute quantum effective actions in the one-loop 
approximation to all orders in both
the coupling and in the external field, so it is well suited to
studying non-perturbative aspects of quantum field theory in strong
background fields. Moreover, the technique maintains gauge invariance
and Lorentz invariance. The key idea of the technique is that a path
integral is approximated as the average of a finite set of $N_l$
representative closed paths (loops) through spacetime. These loops 
are not mapped to any spacetime lattice, so the theory maintains 
Lorentz invariance and is distinct from Lattice-based techniques. We use a
standard Monte-Carlo procedure to generate loops which have large
contributions to the loop average.

We will focus on calculations of the \ac{QED}
effective action in cylindrically symmetric, extended tubes of magnetic 
flux using the worldline numerics method.  These
configurations may be called flux tubes, strings, or vortices,
depending on the context.  Flux tubes are of interest in astrophysics
because they describe magnetic structures near stars and planets,
cosmic strings~\cite{vilenkin2000cosmic}, and vortices in the
superconducting core of neutron stars \cite{2006pfsb.book..135S,
  schmitt2010dense}.  Outside of astrophysics, magnetic vortex systems
are at the forefront of research in condensed matter physics for the
role they play in superconducting systems and in \ac{QCD} research for
their relation to center vortices, a gluonic configuration analogous
to magnetic vortices which is believed to be important to quark
confinement~\cite{tHooft19781, 2003PrPNP..51....1G}.  Currently,
we are most interested in the roles played by magnetic flux tubes in 
neutron star cores.

Our motivation for discussing flux tubes comes from the fact that
superconductivity is predicted in the nuclear matter of neutron stars
and that some superconducting materials produce a lattice of flux
tubes when placed in an external magnetic field. Superconductivity is
a macroscopic quantum state of a fluid of fermions that, most notably,
allows for the resistanceless conduction of charge. In 1933, Meissner
and Ochsenfeld observed that magnetic fields are repelled from
superconducting materials~\cite{meisner33}.  In 1935, F. and H. London
described the Meissner effect in terms of a minimization of the free
energy of the superconducting current~\cite{london35}.  Then, in 1957,
by studying the superconducting electromagnetic equations of motion in
cylindrical coordinates, Abrikosov predicted the possible existence of
line defects in superconductors which can carry quantized magnetic
flux through the superconducting material \cite{abrikosov57}.
A more complete microscopic description of superconducting materials
is given by BCS (Bardeen, Cooper, and Schrieffer)
theory~\cite{PhysRev.108.1175}.  
Interested readers may pursue more thorough
reviews of superconductivity and superfluidity in neutron
stars~\cite{2006pfsb.book..135S, schmitt2010dense}.

Our calculations use the worldline numerics method which is reviewed
in detail in section~\ref{sec:qed-effect-act}.  In section
\ref{sec:background} of this article, we will briefly review the
physics of magnetic flux tubes and of nuclear superconductivity in
neutron stars to provide context and motivation. We present our models
for solitary flux tubes and dense flux tube lattices in section
\ref{sec:calculations} as well as the details of the calculations.
The results of our calculations for scalar and spinor \ac{QED} are
presented in section \ref{sec:periodic_results}. Our results suggest a
small but possibly influential Casimir interaction between flux tubes
in a dense lattice that may cause the flux tubes to form bunches. This
result and other implications of our calculations are discussed in
section \ref{sec:conclusion}.

\section{QED Effective Action on the Worldline}
\label{sec:qed-effect-act}
Worldline numerics is built on the worldline formalism which was
initially invented by Feynman~\cite{PhysRev.80.440, PhysRev.84.108}.
Much of the recent interest in this formalism is based on the work of
Bern and Kosower, who derived it from the infinite string-tension
limit of string theory and demonstrated that it provided an efficient
means for computing amplitudes in QCD~\cite{PhysRevLett.66.1669}.
For this reason, the worldline formalism is often referred to as
`string inspired'. However, the formalism can also be obtained
straight-forwardly from first-quantized field
theory~\cite{1992NuPhB.385..145S}, which is the approach we will adopt
here.  In this formalism the degrees of freedom of the field are
represented in terms of one-dimensional path integrals over an
ensemble of closed trajectories.

\begin{widetext}
We begin with the QED effective action expressed in the proper-time 
formalism \cite{Schwinger:1951},
\be
	\label{eqn:trln}
	\mathrm{Tr~ln}\left[\frac{\slashed{p}+e\slashed{A}_\mu^0
	-m}{\slashed{p}-m}\right] = -\frac{1}{2}\int d^4x \int_0^\infty
	\frac{dT}{T}e^{-iTm^2} 
	 \times \mathrm{tr}\biggr( \bra{x}e^{iT(\slashed{p}
	+e\slashed{A}^0_\mu)^2}\ket{x} 
	- \bra{x}e^{iTp^2}\ket{x}\biggr).
\ee
To evaluate $\bra{x}e^{iT(\slashed{p}_\mu +
  e\slashed{A}_\mu)^2}\ket{x}$, we recognize that it is simply the
propagation amplitude $\braket{x,T}{x,0}$ from ordinary quantum
mechanics with $(\slashed{p}_\mu + e\slashed{A}_\mu)^2$ playing the
role of the Hamiltonian.  We therefore express this factor in its path
integral form:
\be
	\bra{x}e^{iT(\slashed{p}_\mu + e\slashed{A}_mu)^2}\ket{x} =  \mathcal{N}
	\int \mathcal{D}x_\rho(\tau) e^{-\int_0^T d\tau \left[\frac{\dot{x}^2(\tau)}{4} 
	+ i A_\rho x^\rho(\tau)\right]} 
	\times \Paulispin.
\ee
$\mathcal{N}$ is a normalization constant that we can fix by using
our renormalization condition that the fermion determinant should
vanish at zero external field:
\be
	\bra{x}e^{iTp^2}\ket{x}  =  \mathcal{N}\int \mathcal{D} 
		x_p(\tau)e^{-\int_0^T d\tau\frac{\dot{x}^2(\tau)}{4}} 
	= \int \frac{d^4p}{(2\pi T)^4}\bra{x}e^{iTp^2}\ket{p}\braket{p}{x} 
	=  \frac{1}{(4\pi T)^2},	
\ee
We may now write
\be
	\mathcal{N}\int \mathcal{D}x_\rho(\tau) e^{-\int_0^T d\tau[\frac{\dot{x}^2(\tau)}{4}+iA_\rho x^\rho(\tau)]}
	\Paulispin
	= \frac{\left\langle e^{-i\int_0^T d\tau A_\rho x^\rho(\tau)}\Paulispin\right\rangle_x}{(4\pi T)^2} ,
\ee
where
\be
	\label{eqn:meandef}
	\mean{\hat{\mathcal{O}}}_x = \frac{\int \mathcal{D}x_\rho(\tau) \hat{\mathcal{O}} 
	e^{-\int_0^T d\tau\frac{\dot{x}^2(\tau)}{4}}}{\int \mathcal{D}x_\rho(\tau)  
	e^{-\int_0^T d\tau\frac{\dot{x}^2(\tau)}{4}}}
\ee
is the weighted average of the operator $\hat{\mathcal{O}}$ over an ensemble of closed particle 
loops with a Gaussian velocity distribution.

Finally, combining all of the equations from this section results in the 
renormalized one-loop effective action for \ac{QED} on the worldline:
\be
	\label{eqn:QEDWL}
	\Gamma^{(1)}[A_\mu] = \frac{2}{(4\pi)^2}\int_0^\infty
	\frac{dT}{T^3}e^{-m^2T}\int d^4x_\mathrm{CM} \times
        \left[\left\langle e^{i\int_0^Td\tau A_\rho(x_\mathrm{CM}+x(\tau))\dot{x}^\rho(\tau)}
	\Paulispin\right\rangle _x -1\right]. 
\ee
\end{widetext}
\subsection{Worldline Numerics}

The averages, $\mean{\hat{\mathcal{O}}}$, defined by equation (\ref{eqn:meandef})
involve 
functional integration over every possible closed path through spacetime.  The velocities along the paths are drawn from a Gaussian velocity distribution.  
The prescription of the worldline numerics technique is to compute 
these averages approximately using a finite set of $N_l$ representative loops 
on a computer.  The worldline average is then approximated as the mean of 
an operator evaluated along each of the worldlines in the ensemble:
\be
	\mean{\hat{\mathcal{O}}[x(\tau)]} \approx 
	\frac{1}{N_l} \sum_{i=1}^{N_l} \hat{\mathcal{O}}[x_i(\tau)].
\ee

\subsubsection{Loop Generation} 
\label{sec:loopgen}
The velocity distribution for the loops depends on
the proper time, $T$.  However, generating a separate ensemble of loops for 
each value of $T$ would be very computationally expensive.  This problem is alleviated by generating
a single ensemble of loops, $\vec{y}(\tau)$, representing unit proper time,
and scaling those loops accordingly for different values of $T$:
\be 
\vec{x}(\tau) = \sqrt{T}\vec{y}(\tau/T) ,
\ee
\be \int_0^T d\tau \vec{\dot{x}}^2(\tau) \rightarrow \int_0^1 dt
\vec{\dot{y}}^2(t).
\ee
There is no way to treat the integrals as continuous as we generate our loop
ensembles.  Instead, we treat the integrals as sums over discrete points
along the proper-time interval $[0,T]$.  This is fundamentally different
from space-time discretization, however.  Any point on the worldline loop
may exist at any position in space, and $T$ may take on any value.  It is
important to note this distinction because the worldline method retains
Lorentz invariance while lattice techniques, in general, do not.

\begin{figure}
	\centering
		\includegraphics[width=\linewidth,clip,trim=2cm 1cm 2cm 1cm]{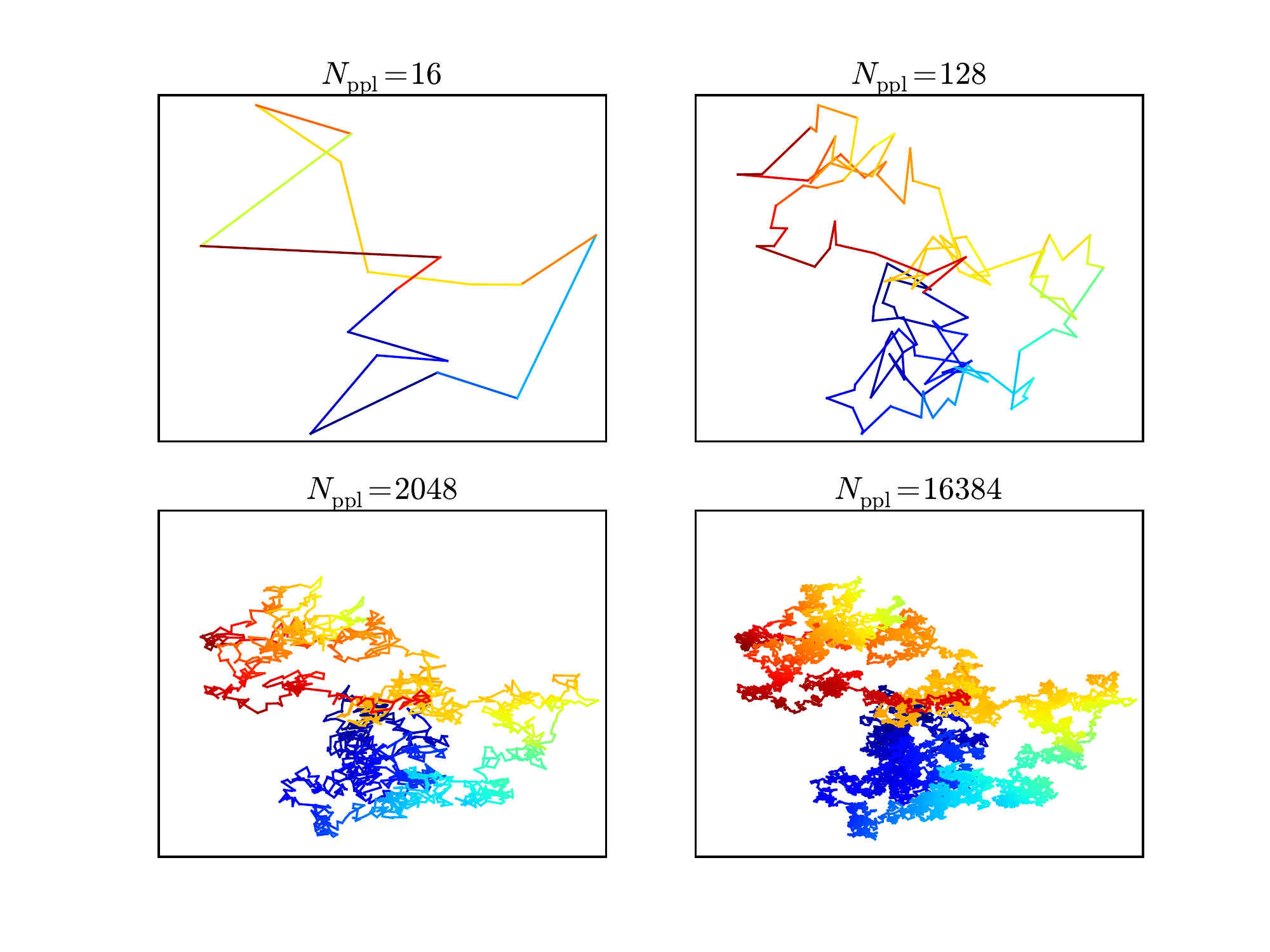}
	\caption[Discretization of the worldline loop]
	{A single discrete worldline loop shown at several levels 
	of discretization. The loops form fractal patterns and have a strong 
	parallel with Brownian motion. The colour
	represents the phase of a particle travelling along the loop, and 
	begins at dark blue, progresses in a random walk through yellow, 
	and ends at dark 
	red. The total flux through this particular worldline at $T=1$ and 
	$B=B_k$ is about $0.08 \pi/e$.}
	\label{fig:worldlineplot}
\end{figure}

The challenge of loop cloud generation is in generating a discrete set
of points on a unit loop which obeys the prescribed velocity distribution.
There are a number of different algorithms for achieving this goal that have
been discussed in the literature.  Four possible algorithms are compared
and contrasted in \cite{Gies:2003cv}.  In this work, we chose an algorithm
dubbed ``d-loops", which was first described
in \cite{Gies:2005sb}. To generate a ``d-loop", the number of points is iteratively 
doubled, placing the new points in a Gaussian distribution between the existing neighbour points.
We quote the algorithm directly:
\begin{enumerate}
\item Begin with one arbitrary point
  $N_0=1$, $y_{N}$.
\item Create an $N_1=2$ loop, i.e., add a point $y_{N/2}$ that is
  distributed in the heat bath of $y_N$ with
  \begin{equation} 
    e^{-\frac{N_1}{4} 2 (y_{N/2} -y_{N})^2}. \label{yn2}
  \end{equation}
\item Iterate this procedure, creating an $N_k=2^k$ points
  per line (ppl) loop by adding $2^{k-1}$ points
  $y_{{qN}/{N_k}}$, $q=1,3,\dots, N_k-1$ with distribution
  \begin{equation} 
    e^{-\frac{N_k}{4} 2 [y_{qN/N_k} -\frac{1}{2}(y_{(q+1)N/N_k}+
	y_{(q-1)N/N_k})]^2}. \label{ynk}
  \end{equation}
\item Terminate the procedure if $N_k$ has reached $N_k=N$ for
  unit loops with $N$ ppl.
\item For an ensemble with common center of mass, shift each
  whole loop accordingly.
\end{enumerate}
The above d-loop algorithm was selected since it is simple and about
10\% faster than previous algorithms, according to its developers,
because it requires fewer algebraic operations.  The generation of the
loops is largely independent from the main program.  Because of this,
it was simpler to generate the loops using a Matlab script.
This function was used to produce text files containing the worldline data for 
ensembles of loops.  These text files were read into memory at the 
launch of each calculation. The results of this generation routine can 
be seen in figure \ref{fig:worldlineplot}.

When the \ac{CUDA} kernel is called, every thread
in every block executes the kernel function with its own unique
identifier. Therefore, it is best to generate worldlines in integer
multiples of the number of threads per block. 

\subsection{Cylindrical Worldline Numerics} 
We now consider cylindrically symmetric external magnetic fields.
In this case, we may simplify (\ref{eqn:QEDWL}),
\ba 
  \label{eqn:cylEA}
\frac{\Gamma^{(1)}_\mathrm{ ferm}}{T
L_z} &=& \frac{1}{4\pi} \int_0^\infty \rho_\mathrm{ cm}
  \biggr[ \int_0^\infty \frac{dT}{T^3}e^{-m^2T} \times \\
& & ~~~~~~ \left\{\langle
  W\rangle_{\vec{r}_\mathrm{ cm}} - 1 -\frac{1}{3}(eB_\mathrm{
  cm}T)^2\right\}\biggr]d\rho. \nonumber
\ea 

\subsubsection{Cylindrical Magnetic Fields} 
\label{sec:cylindrical}

We have $\vec{B} =
B(\rho)$\bhattext{z} with
\be \label{eqn:BWLN} B(\rho) = \frac{A_\phi(\rho)}{\rho} +
\frac{dA_\phi(\rho)}{d\rho} \ee if we make the gauge choice that $A_0 =
A_\rho = A_z = 0$.

We begin by considering $\Aphi$ in the form
\be \Aphi = \frac{F}{2\pi \rho}f_\lambda(\rho) \ee so that \be
B_z(\rho)=\frac{F}{2\pi\rho}\frac{df_\lambda(\rho)}{d\rho} \ee and the total
flux is \be \Phi=F(f_\lambda(L_\rho)-f_\lambda(0)).  \ee It is convenient
to express the flux in units of $\frac{2 \pi}{e}$ and define a dimensionless
quantity
\be 
	\mathcal{F}=\frac{e}{2 \pi} F.
\ee







\subsubsection{Wilson Loop}

The quantity inside the angled brackets in equation (\ref{eqn:QEDWL}) is a 
gauge invariant observable called a Wilson loop. We note that the proper time
integral provides a natural path ordering for this operator.
The Wilson loop expectation value is
\be
\label{eqn:wilsonloop}
\langle W\rangle_{\vec{r}_\mathrm{cm}} \!\! = \biggl \langle e^{ie\int_0^T d\tau
\vec{A}(\vec{r}_{\mathrm{ cm}} + \vec{r}(\tau)) \cdot \dot{\vec{r}}}
 \frac{1}{4} \mathrm{ tr}  e^{\frac{e}{2}\int_0^T d\tau
  \sigma_{\mu \nu}F_{\mu \nu}(\vec{r}_{\mathrm{ cm}} + \vec{r}(\tau))}\biggr
  \rangle_{\vec{r}_\mathrm{ cm}} \!\!\!\!\!\!\!\!,
\ee
which we look at as a product between a scalar part ($e^{ie\int_0^T d\tau
\vec{A}(\vec{r}_{\mathrm{ cm}} + \vec{r}(\tau)) \cdot \dot{\vec{r}}}$)
and a fermionic part ($\frac{1}{4} \mathrm{ tr}  e^{\frac{e}{2}\int_0^T d\tau
  \sigma_{\mu \nu}F_{\mu \nu}(\vec{r}_{\mathrm{ cm}} + \vec{r}(\tau))}$).

\subsubsection{Scalar Part}

In a magnetic field, the scalar part is related to the flux through 
the loop, $\Phi_B$, by Stokes theorem:
\ba
 e^{ie\int_0^T d\tau
 \vec{A} \cdot \dot{\vec{r}}} &=&
 e^{ie\oint \vec{A}\cdot d\vec{r}} = e^{ie\int_{\vec{\Sigma}} \vec{\nabla}\times\vec{A} \cdot d\vec{\Sigma}}\\
 & = & e^{ie\int_{\vec{\Sigma}} \vec{B} d\vec{\Sigma}} = e^{ie\Phi_B}.
\ea
Consequently, this factor accounts for the Aharonov-Bohm phase acquired by 
particles in the loop.

The loop discretization results in the following approximation of the
scalar integral:
\be
  \oint \vec{A}(\vec{r})\cdot d\vec{r} =  \sum_{i=1}^{N_\mathrm{ppl}}
  \int_{\vec{r}^i}^{\vec{r}^{i+1}}\vec{A}(\vec{r})\cdot d\vec{r}.
\ee 
Using a linear parameterization of the positions, the line integrals are
\be 
  \int_{\vec{r}^i}^{\vec{r}^{i+1}}\vec{A}(\vec{r})\cdot d\vec{r} =
  \int_0^1dt \vec{A}(\vec{r}(t))\cdot(\vec{r}^{i+1} - \vec{r}^i).
\ee
Using the same gauge choice outlined above ($\vec{A}=A_\phi \hat{\phi}$),
we may write
\be 
	\vec{A}(\vec{r}(t)) = \frac{\mathcal{F}}{e\rho^2} 
	f_\lambda(\rho^2)(-y,x,0),
\ee 
where we have chosen $f_\lambda(\rho^2)$ to depend on $\rho^2$ instead
of $\rho$ to simplify some expressions and to 
avoid taking many costly square roots in the worldline numerics.
We then have
\be \int_{\vec{r}^i}^{\vec{r}^{i+1}}\vec{A}(\vec{r})\cdot
d\vec{r} = \mathcal{F} (x^iy^{i+1}-y^i x^{i+1})\int_0^1
dt\frac{f_\lambda(\rho_i^2(t))}{\rho_i^2(t)}.  \ee The linear interpolation
in Cartesian coordinates gives
\be 
	\label{eqn:rhoi} \rho_i^2(t) = A_i + 2B_it + C_i t^2,
\ee
where
\ba 
	A_i &=& (x^i)^2 + (y^i)^2 \\ 
	B_i&=&x^i(x^{i+1} - x^i) + y^i(y^{i+1}-y^i)\\ 
	C_i &= &(x^{i+1}-x^i)^2 + (y^{i+1}-y^i)^2. 
	\label{eqn:Ci} 
\ea

In performing the integrals along the straight lines connecting
each discretized loop point, we are in danger of violating gauge invariance.
If these integrals can be performed analytically, than gauge invariance
is preserved exactly.  However, in general, we wish to compute these integrals
numerically.  In this case, gauge invariance is no longer guaranteed, but
can be preserved to any numerical precision that's desired.

\subsubsection{Fermion Part} 

For fermions, the Wilson loop is modified by a factor,
\ba 
W^\mathrm{ ferm.} &=& \frac{1}{4}\mathrm{ tr}\left(e^{\frac{1}{2}
	e\int_0^T d\tau \sigma_{\mu \nu}F^{\mu \nu}}\right)\\ 
	&=& \frac{1}{4}\mathrm{ tr}\left(e^{\sigma_{x y} 
	e\int_0^T d\tau B\left(x(\tau)\right)}\right) \\	
	&=& \cosh{\left(e \int_0^T d\tau B\left(x(\tau)\right)\right)}\\
	&=& \cosh{\left(2\mathcal{F}\int_0^T
	d\tau f'_\lambda(\rho^2(\tau))\right)},
	\label{eqn:Wfermfpl}
\ea 
where we have used the relation 
\be 
	eB = 2\mathcal{F}\frac{d f_\lambda(\rho^2)}{d \rho^2} =
	2\mathcal{F}f'_\lambda(\rho^2).  
\ee 

This factor represents an additional contribution to the 
action because of the spin interaction with the magnetic field. 
Classically, for a particle with a magnetic moment $\vec{\mu}$ 
travelling through a magnetic field in a time $T$, the 
action is modified by a term given by
\be
	\Gamma^0_\mathrm{ spin} = \int_0^T \vec{\mu} \cdot \vec{B}(\vec{x}(\tau)) d\tau.
\ee
The magnetic moment is related to the electron spin 
$\vec{\mu} = g\left(\frac{e}{2m}\right)\vec{\sigma}$, 
so we see that the integral in the above quantum fermion factor is 
very closely related to the classical action 
associated with transporting a magnetic moment through a magnetic field:
\be
	\Gamma^0_\mathrm{ spin} = g\left(\frac{e}{2m}\right) \sigma_{x y} \int_0^T B_z(x(\tau))d\tau.
\ee
Qualitatively, we could write
\be 
	W^\mathrm{ ferm} \sim \cosh{\left(\Gamma^0_\mathrm{ spin}\right)}.
\ee

As a possibly useful aside, we may want to express 
the integral in terms of $f_\lambda(\rho^2)$ instead of its derivative.
We can do this by integrating by parts:
\ba 
	\int_0^T d\tau f'_\lambda(\rho^2(\tau))&=& 
	\frac{T}{N_\mathrm{ ppl}}\sum_{i=1}^{N_\mathrm{ ppl}}\int_0^1
        dt f'_\lambda(\rho^2_i(\tau)) \\ 
         & = & 
	\frac{T}{N_\mathrm{ ppl}}\sum_{i=1}^{N_\mathrm{ ppl}}\biggr[
	\frac{f_\lambda(\rho^2_i(t))}{2(B_i+C_it)}\biggr|^{t=1}_{t=0} \nonumber \\
        & & +\frac{C_i}{2} \!\!\int_0^1\!\!\!\!
  \frac{f_\lambda(\rho^2_i(t))}{(B_i+C_i t)^2} dt\biggr] \\ 
  & = &  \frac{T}{N_\mathrm{ ppl}}\sum_{i=1}^{N_\mathrm{
  ppl}}\frac{C_i}{2}\!\!\int_0^1\!\!\!\! \frac{f_\lambda(\rho^2_i(t))}{(B_i+C_i t)^2} dt,
\ea 
with $\rho_i^2(t)$ given by equations (\ref{eqn:rhoi}) to (\ref{eqn:Ci}).
The second equality is obtained from integration-by-parts.  In the third
equality, we use the loop sum to cancel the boundary terms in pairs:
\be 
\label{eqn:Wfermfl}
W^{\mathrm{ ferm.}} = \cosh{\left(\frac{\mathcal{F}T}{N_\mathrm{
ppl}}\sum_{i=1}^{N_{\mathrm{ ppl}}}
  C_i \int_0^1 dt \frac{f_\lambda(\rho^2_i(t))}{(B_i+C_i t)^2}\right)}.
\ee
In most cases, one would use equation (\ref{eqn:Wfermfpl}) to compute the fermion factor 
of the Wilson loop. However, 
equation (\ref{eqn:Wfermfl}) may be useful in cases where $f'_\lambda(\rho^2(\tau))$
is not known or is difficult to compute. 



\subsubsection{Renormalization}

The field strength renormalization counterterms result from the small $T$
behaviour of the worldline integrand.  In the limit where $T$ is very small,
the worldline loops are very localized around their center of mass.  So,
we may approximate their contribution as being that of a constant field
with value $\vec{A}(\vec{r}_{\mathrm{ cm}})$.  Specifically, we require that the field
change slowly on the length scale defined by $\sqrt{T}$. This condition on 
$T$ can be written
\be 
T \ll \left|\frac{m^2}{e B'(\rho^2)}\right| 
= \left| \frac{m^2}{2\mathcal{F}f''_\lambda(\rho^2_\mathrm{ cm})}\right|.
\ee 

When this limit is satisfied, we may use the exact expressions for the
constant field Wilson loops to determine the small $T$ behaviour of the
integrands and the corresponding counterterms.

The Wilson loop averages for constant magnetic fields in scalar and 
fermionic \ac{QED} are
\be
	\mean{W}_\mathrm{ ferm} = eBT\coth{(eBT)}
\ee
and
\be
	\mean{W}_\mathrm{ scal} = \frac{eBT}{\sinh{(eBT)}}.
\ee
\begin{widetext}
Therefore, the integrand for fermionic \ac{QED} in the limit of small $T$ is
\ba \label{eqn:fermI} I_\mathrm{ ferm}(T) &=&
\frac{e^{-m^2T}}{T^3}\left[eB(\vec{r}_{cm})T\coth{(eB(\vec{r}_{cm})T)}
- 1 -\frac{e^2}{3} B^2(\vec{r}_{cm})T^2\right] \nonumber \\ &\approx&-\frac{(eB)^4
T}{45}+\frac{1}{45} (eB)^4 m^2 T^2+\left( \frac{2 (eB)^6}{945}-\frac{(eB)^4
m^4}{90}\right)T^3 
   +\frac{(7 (eB)^4 m^6-4 (eB)^6 m^2)T^4 }{1890}+O(T^5).
\ea
For scalar QED we have
\ba 
\label{eqn:scalI} I_\mathrm{ scal}(T) &=&
\frac{e^{-m^2T}}{T^3}\left[\frac{eB(\vec{r}_{cm})T}{\sinh{(eB(\vec{r}_{cm})T)}}
- 1 +\frac{1}{6} (eB)^2(\vec{r}_{cm})T^2\right] \nonumber \\ &\approx&\frac{7
(eB)^4 T}{360}-\frac{7  (eB)^4 m^2 T^2}{360}+\frac{(147 (eB)^4 m^4-31
(eB)^6)T^3 }{15120} +\frac{ (31 (eB)^6 m^2-49 (eB)^4
m^6)T^4}{15120}+O(T^5). 
\ea
\end{widetext}

Beyond providing the renormalization conditions, these expansions can
be used in the small $T$ regime to avoid a problem with the Wilson
loop uncertainties in this region.  Consider the uncertainty in the
integrand arising from the uncertainty in the Wilson loop:
\be
\delta I(T) = \frac{\partial I}{\partial W} \delta W = \frac{e^{-m^2 T}}{T^3} \delta W.
\ee 
In this case, even though we can compute the Wilson loops for small $T$
precisely, even a small uncertainty is magnified by a divergent factor when
computing the integrand for small values of $T$.  So, in order to perform
the integral, we must replace the small $T$ behaviour of the integrand with
the above expansions (\ref{eqn:fermI}) and (\ref{eqn:scalI}).  Our worldline
integral then proceeds by analytically computing the integral for the small
$T$ expansion up to some small value, $a$, and adding this to the remaining
part of the integral~\cite{MoyaertsLaurent:2004}:
\be 
	\int_0^{\infty} I(T) dT = \underbrace{\int_0^a I(T) dT}_\mathrm{ small ~T} 
	+ \underbrace{\int_a^\infty I(T)dT}_\mathrm{ worldline ~numerics}.
\ee
Because this normalization procedure uses the constant field expressions for small values of 
$T$, this scheme introduces a small systematic uncertainty. To improve on the 
method outlined here, the derivatives of the background field can 
be accounted for by using the analytic forms of the heat kernel expansion to perform the 
renormalization~\cite{Gies:2001tj}. 

\subsection{Uncertainty analysis in worldline numerics} 
\label{ch:WLError}

So far in the worldline numerics literature, the discussions of uncertainty
analysis have been unfortunately brief.  It has been suggested that the
standard deviation of the worldlines provides a good measure of the
statistical error in the worldline method~\cite{Gies:2001zp,
  Gies:2001tj}.  However, the distributions produced by the worldline
ensemble are highly non-Gaussian (see figure \ref{fig:hists}), 
and therefore the standard error in
the mean is not a good measure of the uncertainties
involved. Furthermore, the use of the same worldline ensemble to
compute the Wilson loop multiple times in an integral results in
strongly correlated uncertainties. Thus, propagating uncertainties
through integrals can be computationally expensive due to the
complexity of computing correlation coefficients.

The error bars on worldline calculations impact the conclusions that
can be drawn from calculations, and also have important implications
for the fermion problem, which limits the domain of applicability of
the technique (see section \ref{sec:fermionproblem}).  It is therefore
important that the error analysis is done thoughtfully and
transparently.  The purpose of this section is to contribute a more
thorough discussion of uncertainty analysis in the worldline numerics technique
to the literature in hopes of avoiding any confusion associated with
the above-mentioned subtleties.

There are two sources of uncertainty in the worldline technique: the
discretization error in treating each continuous worldline as a set of
discrete points, and the statistical error of sampling a finite number
of possible worldlines from a distribution.  In this section, we
discuss each of these sources of uncertainty.

\subsubsection{Estimating the Discretization Uncertainties}
\label{sec:discunc}

The discretization error arising from the integral over $\tau$ in the
exponent of each Wilson loop (see equation (\ref{eqn:wilsonloop})) is
difficult to estimate since any number of loops could be represented
by each choice of discrete points.  The general strategy is to make
this estimation by computing the Wilson loop using several different
numbers of points per worldline and observing the convergence
properties.

The specific procedure adopted for this work involves dividing each
discrete worldline into several worldlines with varying levels of
discretization.  Since we are using the d-loop method for generating
the worldlines (section \ref{sec:loopgen}), a
$\frac{N_\mathrm{ppl}}{2}$ sub-loop consisting of every other point
will be guaranteed to contain the prescribed distribution of
velocities.

\begin{figure}
	\centering
		\includegraphics[width=\linewidth]{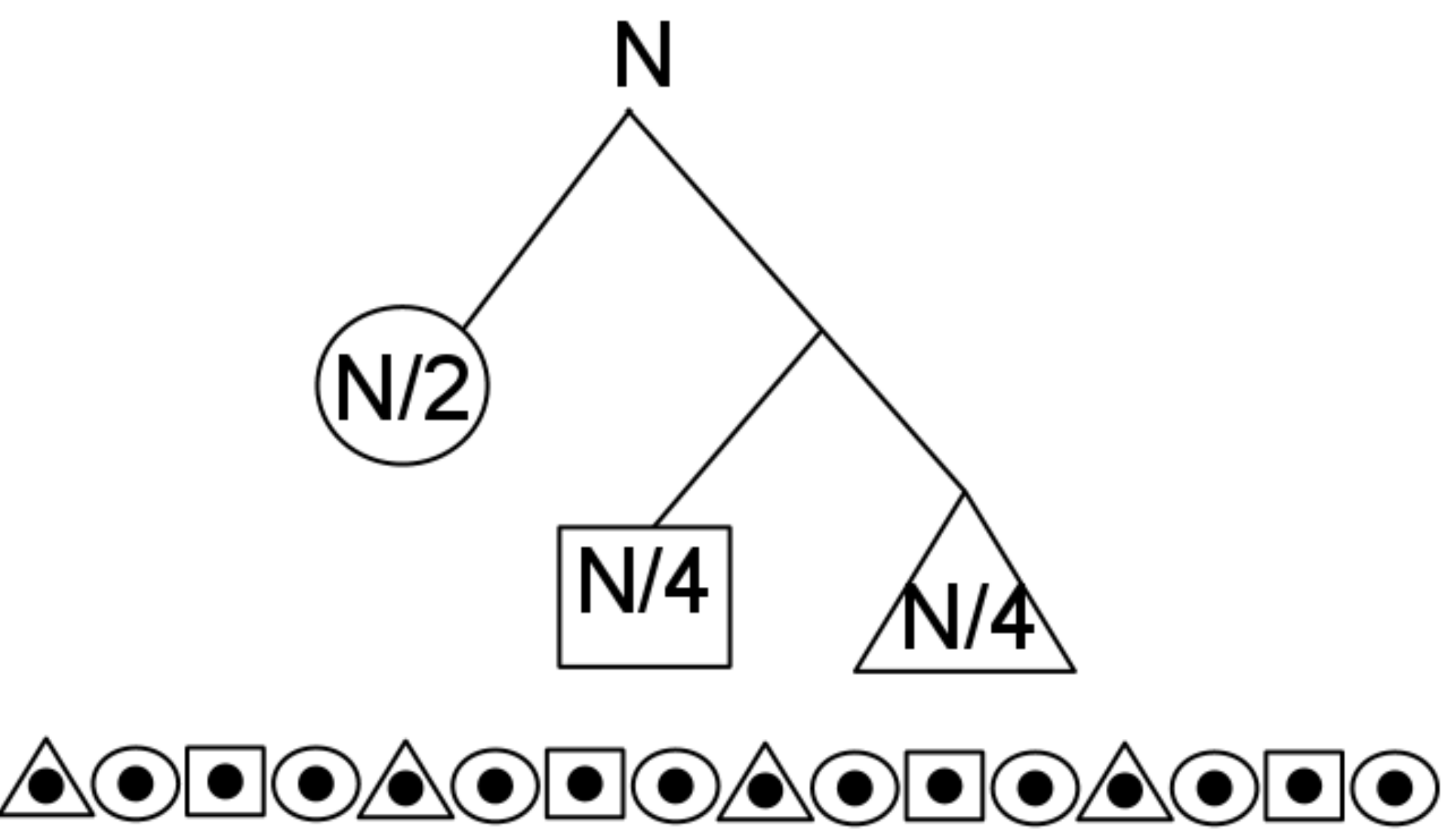}
	\caption[Illustration of convergence testing scheme]
	{Diagram illustrating the division of a worldline into three smaller interleaved worldlines}
	\label{fig:Division}
\end{figure}

To look at the convergence for the loop discretization, 
each worldline is divided into three groups.  One group of $\frac{N_\mathrm{ppl}}{2}$ points, and two groups of 
$\frac{N_\mathrm{ppl}}{4}$.  
This permits us to compute the average holonomy factors at three levels of 
discretization:
\be 
\mean{W}_{N_\mathrm{ppl}/4} = \mean{e^{\frac{i}{2}\triangle} e^{\frac{i}{2}\Box}},
\ee 
\be 
\mean{W}_{N_\mathrm{ppl}/2} = \mean{e^{i\circ}},
\ee 
and
\be 
\mean{W}_{N_\mathrm{ppl}} = \mean{e^{\frac{i}{2}\circ} e^{\frac{i}{4}\Box} e^{\frac{i}{4}\triangle}},
\ee 
where the symbols $\circ$, $\Box$, and $\triangle$ denote the worldline integral, 
$\int_0^T d\tau A(x_{CM}+x(\tau))\cdot \dot x$, computed using the sub-worldlines 
depicted in figure \ref{fig:Division}.

\begin{figure}
	\centering
		\includegraphics[width=\linewidth,clip,trim=1.8cm 0.3cm 1.8cm 1cm]{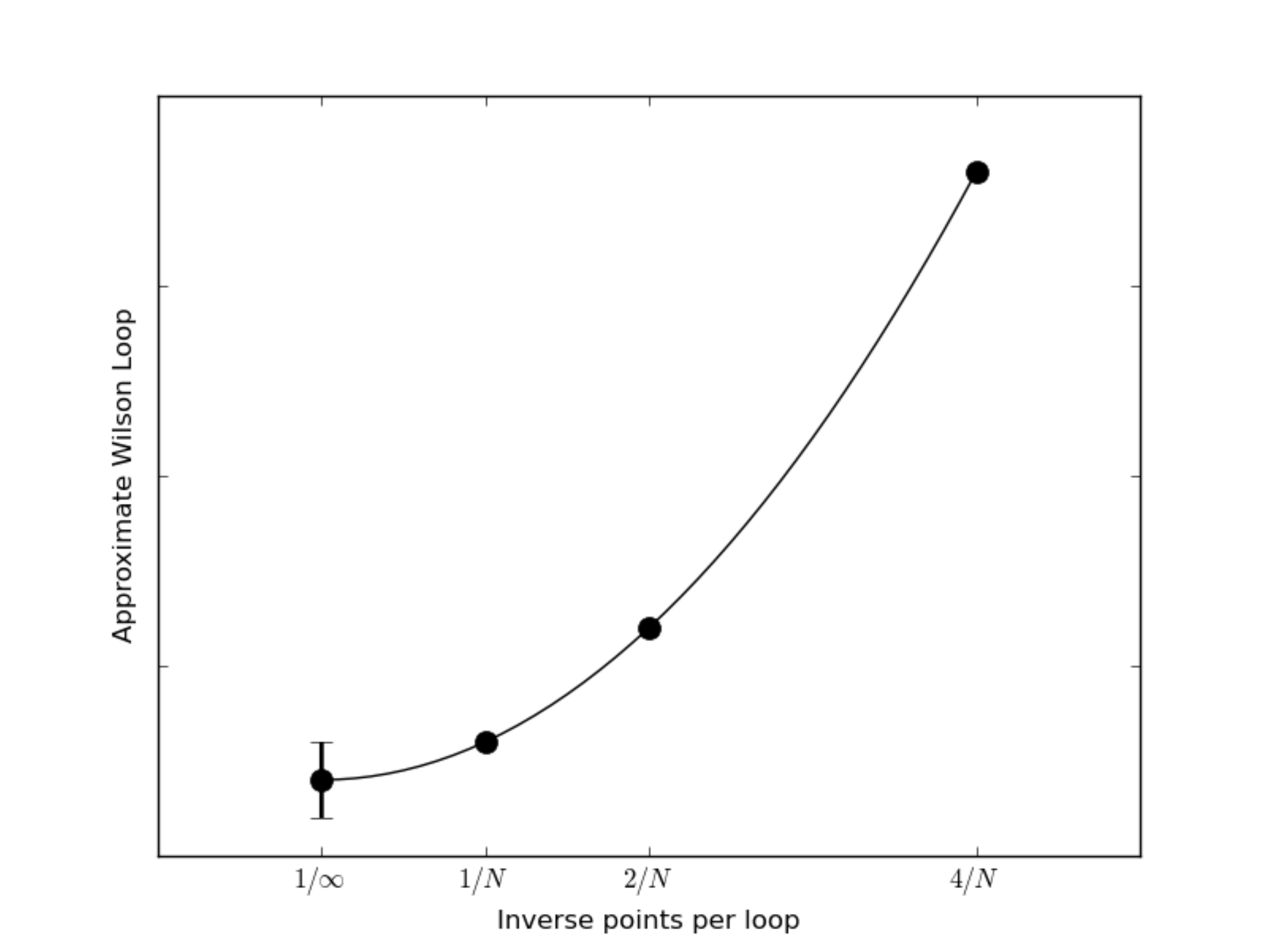}
	\caption[Illustration of extrapolation to infinite points per loop]
	{This plot illustrates the method used to extrapolate the Wilson loop to 
	infinite points per loop and the uncertainty estimate in the approximation.}
	\label{fig:DiscErr}
\end{figure}

We may put these factors into the equation of a parabola to extrapolate the result to an infinite 
number of points per line (see figure \ref{fig:DiscErr}):
\be 
\mean{W}_{\infty} \approx \frac{8}{3}\mean{W}_{N_\mathrm{ppl}} - 2 \mean{W}_{N_\mathrm{ppl}/2} +\frac{1}{3}\mean{W}_{N_\mathrm{ppl}/4}.
\ee 
So, we estimate the discretization uncertainty to be
\be 
\delta \mean{W}_{\infty} \approx |\mean{W}_{N_\mathrm{ppl}} -  \mean{W}_{\infty}|.
\ee 
Generally, the statistical uncertainties are the limitation in the precision of the 
worldline numerics technique. Therefore, $N_\mathrm{ppl}$ should be chosen to be 
large enough that the discretization uncertainties are small relative to the 
statistical uncertainties.

\subsubsection{Estimating the Statistical Uncertainties}

We can gain a great deal of insight into the nature of the statistical uncertainties 
by examining the specific case of the uniform magnetic field since we know the 
exact solution in this case. Sections~\ref{sec:notnormal},~\ref{sec:correlations},
and~\ref{sec:groupingworldlines} discuss the peculiarities of the 
statisical uncertainties in the worldline numerics method for the uniform magnetic
field. 

\subsubsection{The Worldline Ensemble Distribution is not Normal}
\label{sec:notnormal}

\begin{figure}
	\centering
		\includegraphics[width=\linewidth,clip,trim=0.8cm 0.3cm 1.8cm 1cm]{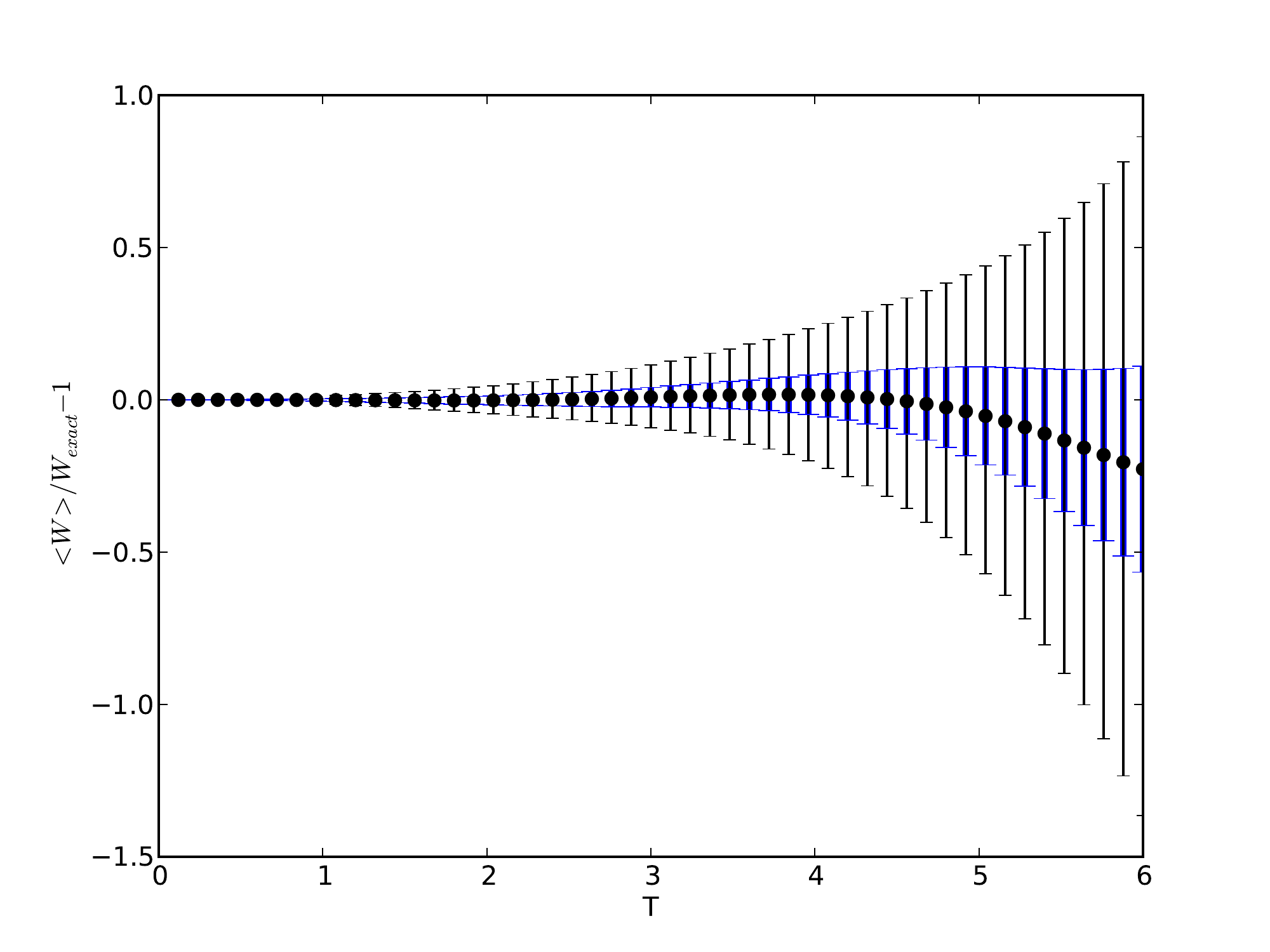}
	\caption[Comparison of error bars between standard error in
          the mean and jackknife analysis] {The residuals of the
          Wilson loops for a constant magnetic field showing the
          standard error in the mean (thin error bars) and the
          uncertainty in determining the mean (thick blue error
          bars). For reasons discussed in this section, the standard
          error in the mean overestimates the uncertainties involved
          by more than a factor of 3 at each value of $T$.}
	\label{fig:resids2}	\label{fig:resids}
\end{figure}

A reasonable first instinct for estimating the error bars is to use the standard 
error in the mean of the collection of individual worldlines:
\be 
\mathrm{ SEM}( W ) = \sqrt{\sum_{i=1}^{N_l}\frac{(W_i - \mean{W})^2}{N_l(N_l-1)}}.
\ee 
This approach has been promoted in early papers on worldline numerics~\cite{Gies:2001zp, Gies:2001tj}.
In figure \ref{fig:resids}, we have plotted the residuals and the corresponding 
error bars for several values of the proper time parameter, $T$ in black.  From this plot, 
it appears that the error bars are quite large in the sense that we appear to 
produce residuals which are considerably smaller than would be implied by the 
sizes of the error bars.  This suggests that we have overestimated the size of 
the uncertainty.

\begin{figure}
	\centering
		\includegraphics[width=\linewidth,clip,trim=0.8cm 0.3cm 1.8cm 1cm]{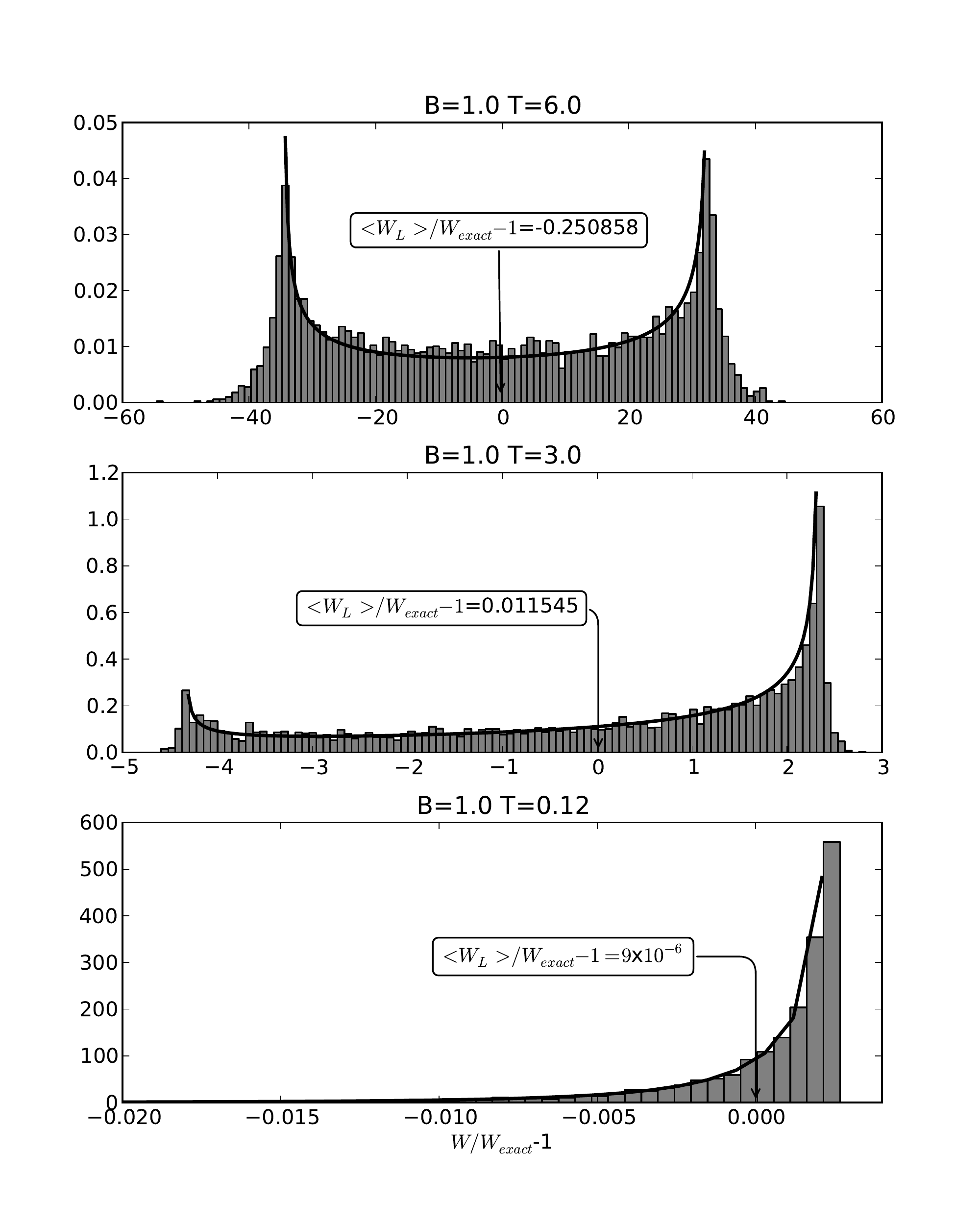}
	\caption[Histograms showing the worldline distributions]
	{Histograms showing the worldline distributions of the residuals 
	for three values of $T$ in the constant magnetic field case. 
	Here, we are neglecting the fermion factor. The dark 
	line represents the exact distribution computed using equation 
	\ref{eqn:exactdist}.  The worldline means are indicated with an arrow, 
	while the exact mean in each case is 0.  There are 5120 worldlines in 
	each histogram.  The vertical axes are normalized to a total area of unity.}
	\label{fig:hists}
\end{figure}

We can see why this is the case by looking more closely at the
distributions produced by the worldline technique.  An exact
expression for these distributions can be derived in the case of the
constant magnetic field~\cite{MoyaertsLaurent:2004}: 
\ba
\label{eqn:exactdist}
w(y)&=&\frac{W_\mathrm{ exact}}{\sqrt{1-y^2}}\sum_{n=-\infty}^{\infty}\biggl[f(\arccos(y)+2n\pi)+\nonumber \\
& & ~~~~~~~~~~~~ f(-\arccos(y)+2n\pi)\biggr]
\ea
with
\be 
f(\phi)=\frac{\pi}{4BT\cosh^2(\frac{\pi \phi}{2BT})}.
\ee 
Figure \ref{fig:hists} shows histograms of the worldline results along with the 
expected distributions.  These distributions highlight a significant hurdle in 
assigning error bars to the results of worldline numerics.

Due to their highly non-Gaussian nature, the standard error in the mean is not a good 
characterization of the distributions that are produced.  We should not interpret each individual 
worldline as an independent measurement of the mean value of these distributions; for large values of $BT$, 
almost all of our worldlines will produce answers which are far away from the mean of the 
distribution. This means that the variance of the distribution will be very large, even 
though our ability to determine the mean of the distribution is relatively precise
because of the increasing symmetry about the mean as $T$ becomes large.

\subsubsection{Correlations between Wilson Loops}
\label{sec:correlations}

\begin{figure}
	\centering
		\includegraphics[width=\linewidth,clip,trim=0.8cm 0.3cm 1.8cm 1cm]{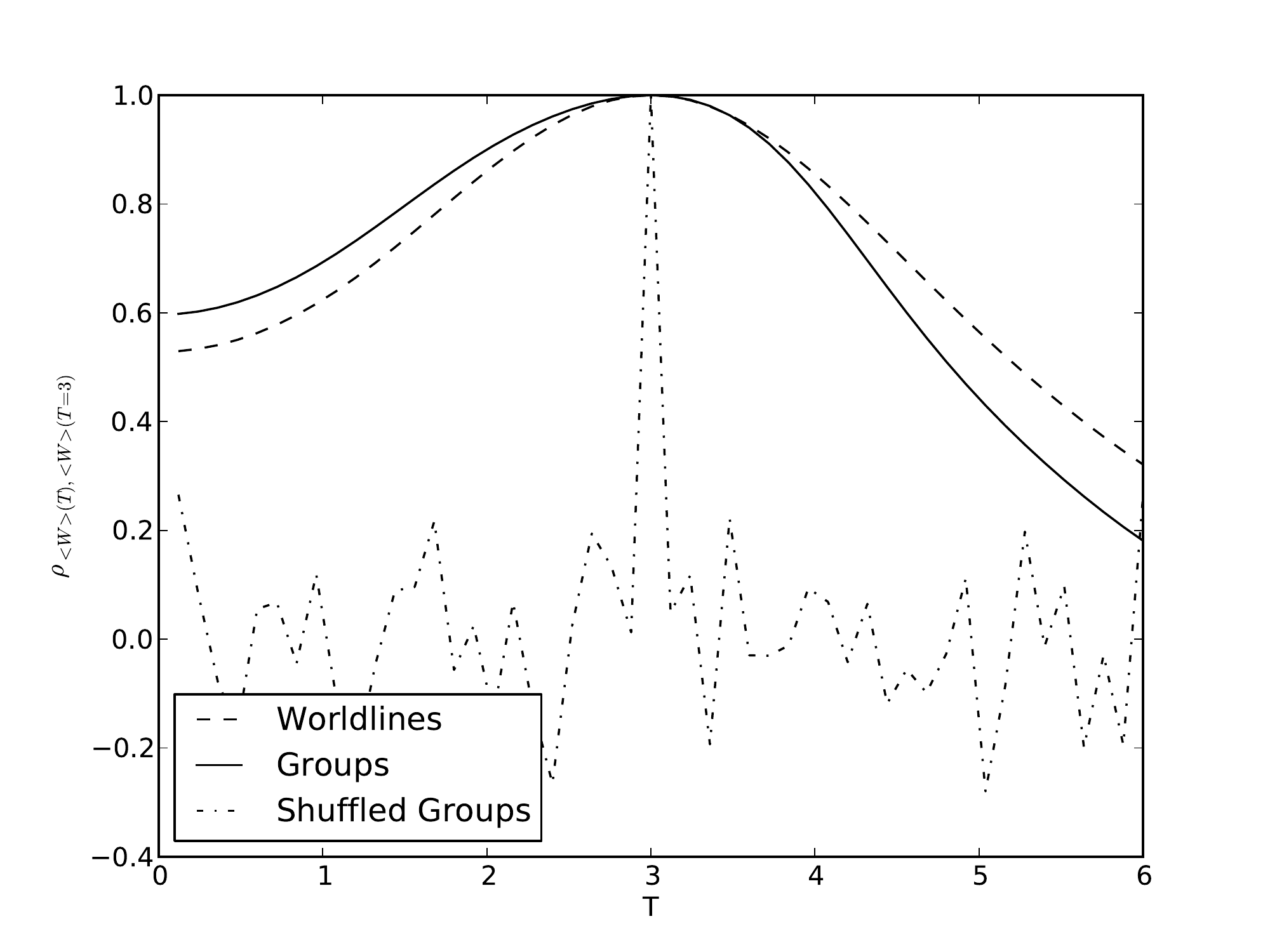}
	\caption[Correlation coefficients between different evaluations of the integrand]
	{Correlation coefficients, equation (\ref{eqn:corrcoef}), between $\mean{W(T)}$ 
		and $\mean{W(T=3)}$ computed using individual worldlines, groups of worldlines, and 
		shuffled groups of worldlines.}
	\label{fig:corr}
\end{figure}

Typically, numerical integration is performed by replacing the integral with a sum over a finite set 
of points from the integrand.  We will begin the present discussion by considering the uncertainty 
in adding together two points (labelled $i$ and $j$) in our integral over $T$. Two terms of the 
sum representing the numerical integral will involve a function of $T$ times the two 
Wilson loop factors,
\be 
I = g(T_i)\mean{W(T_i)} + g(T_j)\mean{W(T_j)}
\ee 
with an uncertainty given by
\ba 
	\delta I &=& \left | \pderiv{I}{\mean{W(T_i)}} \right|^2 (\delta \mean{W(T_i)})^2   \\ \nonumber
	& &
	~~ + \left | \pderiv{I}{\mean{W(T_j)}} \right|^2 (\delta
        \mean{W(T_j)})^2  \\ \nonumber
	& & ~~ + 2 \left | \pderiv{I}{\mean{W(T_i)}}
          \pderiv{I}{\mean{W(T_j)}} \right | \times \\ \nonumber 
& & ~~\rho_{ij} 
	(\delta \mean{W(T_i)}) (\delta \mean{W(T_j)}) \\
	&=& g(T_i)^2 (\delta \mean{W(T_i)})^2 + g(T_j)^2 (\delta \mean{W(T_j)})^2+ \nonumber \\
	& & ~ 2 \left | g(T_i)g(T_j) \right | \rho_{ij} (\delta \mean{W(T_i)}) (\delta \mean{W(T_j)})
\ea 
and the correlation coefficient $\rho_{ij}$ given by
\be 
	\label{eqn:corrcoef}
	\rho_{ij} = \frac{\mean{(W(T_i) - \mean{W(T_i)})(W(T_j)-\mean{W(T_j)})}}{\sqrt{(W(T_i)
	-\mean{W(T_i)})^2}\sqrt{(W(T_j)-\mean{W(T_j)})^2}}.
\ee 
The final term in the error propagation equation takes into account correlations between the 
random variables $W(T_i)$ and $W(T_j)$. Often in a Monte Carlo computation, one can 
treat each evaluation of the integrand as independent, and neglect the uncertainty
term involving the correlation coefficient. However, in worldline numerics, 
the evaluations are related because the same worldline ensemble is reused 
for each evaluation of the integrand.
The correlations are significant (see figure \ref{fig:corr}), and this term 
can't be neglected. Computing each correlation coefficient takes 
a time proportional to the square of the number of worldlines. Therefore, it may 
be computationally expensive to formally propagate uncertainties through an 
integral.

The point-to-point correlations were originally pointed out by Gies
and Langfeld who addressed the problem by updating (but not replacing
or regenerating) the loop ensemble in between each evaluation of the
Wilson loop average~\cite{Gies:2001zp}.  This may be a good way of
addressing the problem. However, in the following section, we promote a
method which can bypass the difficulties presented by the correlations
by treating the worldlines as a collection of worldline groups.

\subsubsection{Grouping Worldlines}
\label{sec:groupingworldlines}

Both of the problems explained in the previous two subsections can be overcome 
by creating groups of worldline loops within the ensemble. Each group of worldlines 
then makes a statistically independent measurement of the Wilson loop average 
for that group. The statistics between the groups of measurements are normally
distributed, and so the uncertainty is the standard error in the mean of the 
ensemble of groups (in contrast to the ensemble of worldlines).

For example, if we divide the $N_l$ worldlines into $N_G$ groups of $N_l/N_G$ 
worldlines each, we can compute a mean for each group:
\be
	\mean{W}_{G_j} = \frac{N_G}{N_l}\sum_{i=1}^{N_l/N_G}W_i.
\ee
Provided each group contains the same number of worldlines, 
the average of the Wilson loop is unaffected by this grouping:
\ba
	\mean{W} & = & \frac{1}{N_G} \sum_{j=1}^{N_G} \mean{W}_{G_j} \\
		& = & \frac{1}{N_l} \sum_{i=1}^{N_l} W_i.
\ea
However, the uncertainty is the standard error in the mean of 
the groups,
\be
	\delta \mean{W} = \sqrt{\sum_{i=1}^{N_G} 
		\frac{(\mean{W}_{G_i} - \mean{W})^2}{N_G(N_G-1)}}.
\ee

Because the worldlines are unrelated to one another, the choice of how to 
group them to compute a particular Wilson loop average is arbitrary. For example, 
the simplest choice is to group the loops by the order they were generated, so that 
a particular group number, $i$, contains worldlines $iN_l/N_G$ through $(i+1)N_l/N_G -1$. 
Of course, if the same worldline groupings are used to compute different Wilson 
loop averages, they will still be correlated. We will discuss this problem in a moment.

\begin{figure}
	\centering
		\includegraphics[width=\linewidth,clip,trim=0.8cm 0.3cm 1.8cm 1cm]{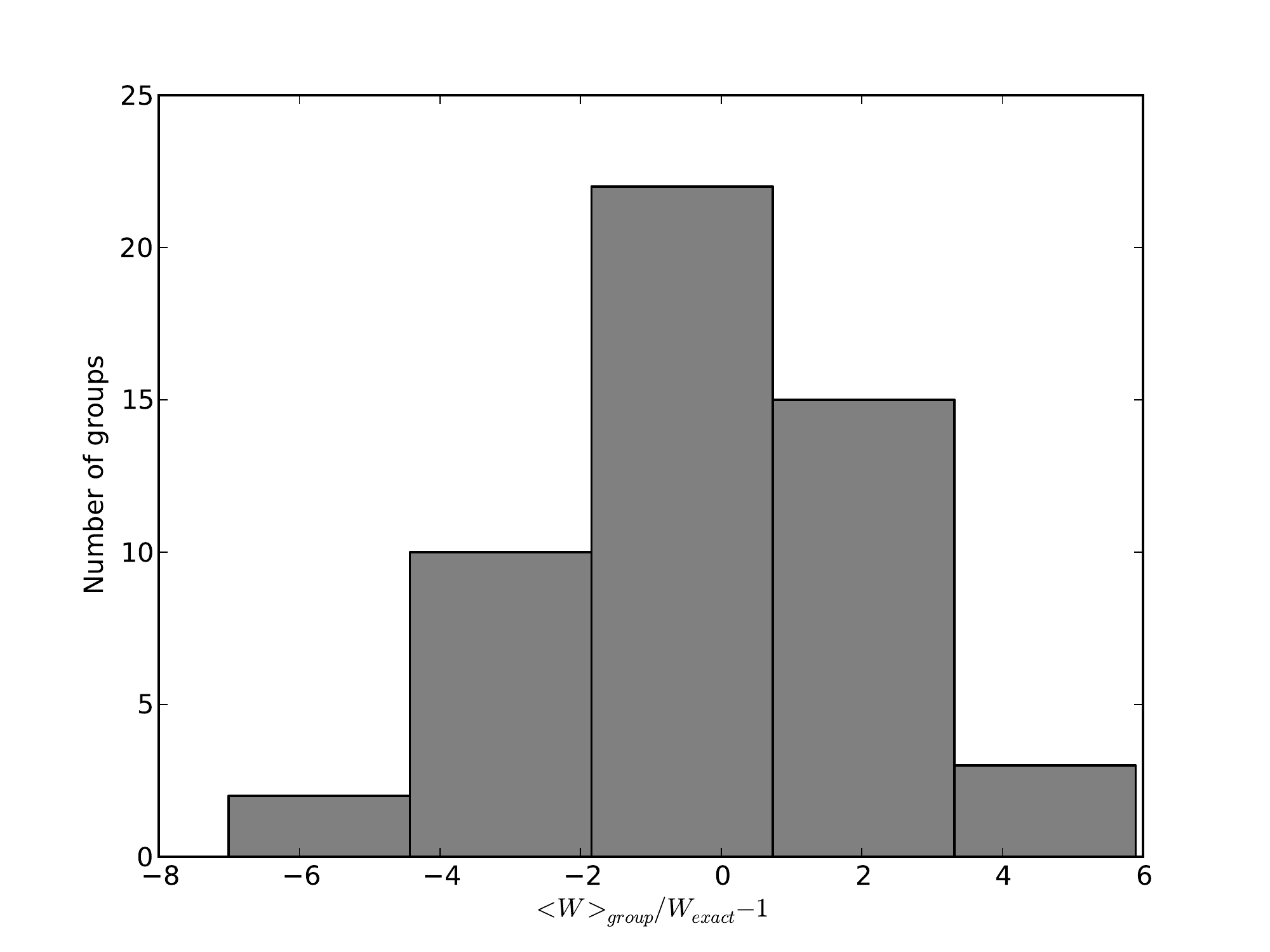}
	\caption[Histogram for reproducing measurements with groups of worldlines]
	{The histogram demonstrating the precision with which we can reproduce 
		measurements of the mean using different groups of 100 worldlines at $BT=6.0$.  
		In this case, the distribution is Gaussian-like and meaningful error bars can 
		be placed on our measurement of the mean.}
	\label{fig:uncinmean}
\end{figure}

The basic claim of the worldline technique is that the mean of the
worldline distribution approximates the holonomy factor. However, from
the distributions in figure \ref{fig:hists}, we can see that the
individual worldlines themselves do not approximate the holonomy
factor. So, we should not think of an individual worldline as an
estimator of the mean of the distribution.  Thus, a resampling
technique is required to determine the precision of our statistics. We
can think of each group of worldlines as making an independent
measurement of the mean of a distribution. As expected, the groups of
worldlines produce a more Gaussian-like distribution (see figure
\ref{fig:uncinmean}), and so the standard error of the groups is a
sensible measure of the uncertainty in the Wilson loop value.

We find that the error bars are about one-third as large as those
determined from the standard error in the mean of the individual
worldlines, and the smaller error bars better characterize the size of
the residuals in the constant field case (see figure
\ref{fig:resids2}).  The strategy of using subsets of the available
data to determine error bars is called jackknifing.  Several previous
papers on worldline numerics have mentioned using jackknife analysis to
determine the uncertainties, but without an explanation of the
motivations or the procedure employed \cite{2005PhRvD..72f5001G,
  PhysRevLett.96.220401, Dunne:2009zz, PhysRevD.84.065035}.

The grouping of worldlines alone does not address the problem of
correlations between different evaluations of the integrands.  Figure
\ref{fig:corr} shows that the uncertainties for groups of worldlines
are also correlated between different points of the integrand.
However, the worldline grouping does provide a tool for bypassing the
problem. One possible strategy is to randomize how worldlines are
assigned to groups between each evaluation of the integrand. This
produces a considerable reduction in the correlations, as is shown in
figure \ref{fig:corr}. Then, errors can be propagated through the
integrals by neglecting the correlation terms. Another strategy is to
separately compute the integrals for each group of worldlines, and
then consider the statistics of the final product to determine the
error bars. This second strategy is the one adopted for the work
presented in this paper. Grouping in this way reduces the amount of
data which must be propagated through the integrals by a factor of the
group size compared to a delete-1 jackknife scheme, for example.  In
general, the error bars quoted in the remainder of this paper are obtained by
computing the standard error in the mean of groups of worldlines.

\subsubsection{Uncertainties and the Fermion Problem}
\label{sec:fermionproblem}

The fermion problem of worldline numerics is a name given to an enhancement of
the uncertainties at large $T$~\cite{Gies:2001zp,
  MoyaertsLaurent:2004}. It should not be confused with the
fermion-doubling problem associated with lattice methods. In a
constant magnetic field, the scalar portion of the calculation
produces a factor of $\frac{BT}{\sinh{(BT)}}$, while the fermion
portion of the calculation produces an additional factor
$\cosh{(BT)}$. Physically, this contribution arises as a result of the
energy required to transport the electron's magnetic moment around the
worldline loop.  At large values of $T$, we require subtle
cancellation between huge values produced by the fermion portion with
tiny values produced by the scalar portion.  However, for large $T$,
the scalar portion acquires large relative uncertainties which make
the computation of large $T$ contributions to the integral very
imprecise.

This can be easily understood by examining the worldline distributions shown in figure
\ref{fig:hists}. Recall that the scalar Wilson loop average for these histograms is given 
by the flux in the loop, $\Phi_B$:
\be
	\mean{W} = \left<\exp{\left(ie\int_0^Td\tau \vec{A}(\vec{x}_\mathrm{ cm} 
	+ \vec{x}(\tau))\cdot d\vec{x}(\tau)\right)}\right> = \left< e^{ie\Phi_B}\right>.
\ee
For constant fields, the flux through the worldline loops obeys the distribution function
\cite{MoyaertsLaurent:2004}
\be
	f(\Phi_B) = \frac{\pi}{4BT\cosh^2\left(\frac{\pi \Phi_B}{2BT}\right)}.
\ee
For small values of $T$, the worldline loops are small and the 
amount of flux through the loop is correspondingly small. Therefore, the 
flux for small loops is narrowly distributed about $\Phi_B = 0$. Since zero 
maximizes the Wilson loop ($e^{i0}=1$), 
this explains the enhancement to the right of the distribution for small values of $T$. 
As $T$ is increased, the flux through any given worldline becomes very large and the 
distribution of the flux becomes very broad. 
For very large $T$,  the width of the distribution is many 
factors of $2\pi/e$. Then, the phase ($e \Phi_B\mod{2\pi}$) is nearly 
uniformly distributed, and the 
Wilson loop distribution reproduces the Chebyshev distribution (\ie 
the distribution obtained from projecting uniformly distributed points on 
the unit circle onto the horizontal axis),
\be
	\lim_{T\to\infty}w(y) = \frac{1}{\pi\sqrt{1-y^2}}.
\ee

The mean of the Chebyshev distribution is zero due to its symmetry. 
However, this symmetry is not 
realized precisely unless we use a very large number of loops. Since the 
width of the distribution is already 100$\times$ the value of the mean at 
$T=6$, any numerical asymmetries in the distribution result in very large 
relative uncertainties of the scalar portion. Because of these uncertainties, 
the large contribution from the fermion factor are not cancelled precisely.

This problem makes it very difficult to compute 
the fermionic effective action unless the fields are well localized
\cite{MoyaertsLaurent:2004}. For example, the fermionic factor for 
non-homogeneous magnetic fields oriented along the z-direction is
\be
	\cosh{\left(e\int_0^T d\tau B(x(\tau))\right)}.
\ee
For a homogeneous field, this function grows exponentially with $T$ and 
is cancelled by the exponentially vanishing scalar Wilson loop.
For a localized field, 
the worldline loops are very large for large values of $T$, and they primarily 
explore regions far from the field. Thus, the fermionic factor grows more slowly 
in localized fields, and is more easily cancelled by the rapidly vanishing scalar part.


\begin{figure}[ht] 
  \begin{center}
    \includegraphics[width=\linewidth,clip,trim=0.8cm 0.3cm 1.8cm 1cm]{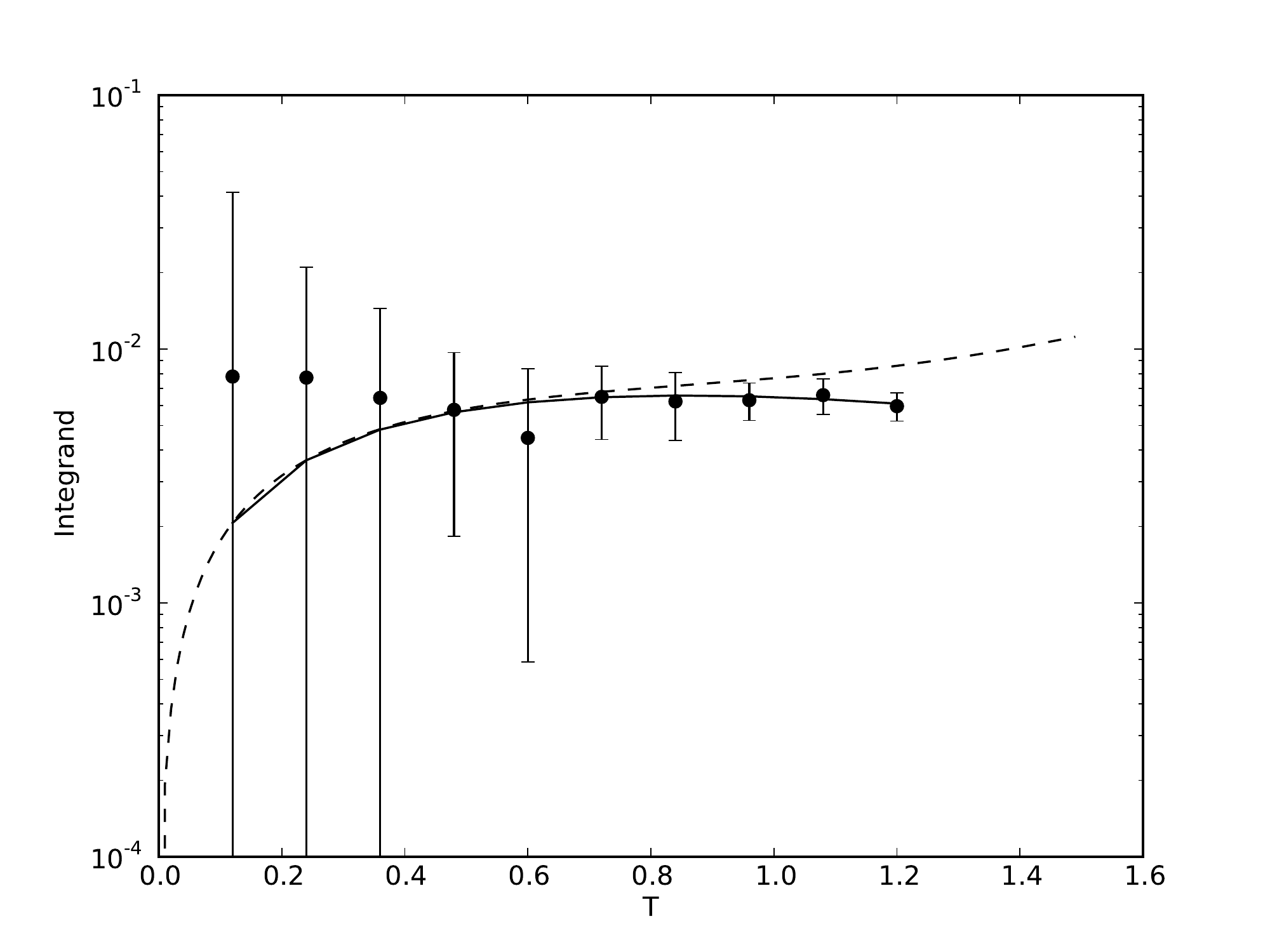} 
	    \caption[small $T$ behaviour of worldline numerics]
		{The small $T$
		behaviour of worldline numerics.  The data points represent the numerical
		results, where the error bars are determined from the jackknife analysis described 
		in chapter \ref{ch:WLError}.
		The solid line represents the exact solution while the dotted line represents
		the small $T$ expansion of the exact solution.  Note the amplification of
		the uncertainties.} 
  \end{center} 
\end{figure}

\subsection{Computing an Effective Action}

The ensemble average in the effective action is simply the sum over the
contributions from each worldline loop, divided by the number of loops in
the ensemble.  Since the computation of each loop is independent of the
other loops, the ensemble average may be straightforwardly parallelized by
generating separate processes to compute the contribution from each loop.
For this parallelization, four Nvidia Tesla C1060 \acp{GPU} were used through
the \ac{CUDA} parallel programming framework.  Because \acp{GPU} can spawn
thousands of parallel processing threads
with much less computational overhead
than an \ac{MPI} cluster, they excel at handling a very large number of parallel 
threads, although the clock speed is slower and fewer memory resources are typically available.
The \ac{GPU} architecture has 
recently been used by another group for computing Casimir forces using 
worldline numerics~\cite{2011arXiv1110.5936A}. More detailed information about the 
technical implementation of these calculations on the \ac{GPU} architecture, 
including a listing of the source code,  can be 
found in~\cite{2012PhDT........21M}.

Once the ensemble average of the Wilson loop has been computed,
computing the effective action is a straightforward matter of
performing numerical integrals.  The effective action density is
computed by performing the integration over proper time, $T$.  Then,
the effective action is computed by performing a spacetime integral
over the loop ensemble center of mass.  In all cases where a numerical
integral was performed, Simpson's method was
used~\cite{burden2001numerical}. Integrals from 0 to $\infty$ were
mapped to the interval $[0,1]$ using substitutions of the form $x =
\frac{1}{1+T/T_\mathrm{ max}}$, where $T_\mathrm{ max}$ sets the scale
for the peak of the integrand. In the constant field case, for the
integral over proper time, we expect $T_\mathrm{ max} \sim 3/(eB)$ for
large fields and $T_\mathrm{ max} \sim 1$ for fields of a few times
critical or smaller. In section~\ref{ch:WLError}, we presented a
detailed discussion of how the statistical and discretization
uncertainties can be computed in this technique.



\subsection{Verification and Validation}
\label{sec:verify}

The worldline numerics software can be validated and verified by making sure that it 
produces the correct results where the derivative expansion is a good approximation, 
and that the results are consistent with previous numerical calculations of flux tube 
effective actions. For this reason, the validation was done primarily with flux tubes with 
a profile defined by the function
\be
	f_\lambda(\rho^2) = \frac{\rho^2}{(\lambda^2 + \rho^2)}.
\ee
For large values of $\lambda$, this function varies slowly on the Compton wavelength 
scale, and so the derivative expansion is a good approximation. Also, flux tubes 
with this profile were studied previously using worldline numerics
\cite{Moyaerts:2003ts, MoyaertsLaurent:2004}.

Among the results presented in~\cite{MoyaertsLaurent:2004} is a comparison of 
the derivative expansion and worldline numerics for this magnetic field 
configuration. The result is that the next-to-leading-order term 
in the derivative expansion is only a small correction to the the 
leading-order term for $\lambda \gg \lambda_e$, where the derivative 
expansion is a good approximation. The derivative expansion breaks
down before it reaches its formal validity limits 
at $\lambda \sim \lambda_e$. For this reason,
we will simply focus on the leading order derivative expansion, which we call 
the locally-constant-field (\ac{LCF}) approximation.
The effective action of \ac{QED} in the \ac{LCF} approximation is given 
in cylindrical symmetry by
\ba
	\label{eqn:LCFferm}
	\Gamma^{(1)}_\mathrm{ ferm} &=& \frac{1}{4\pi}\int_0^\infty dT 
	\int_0^\infty \rho_\mathrm{ cm} d\rho_\mathrm{ cm}\frac{e^{-m^2T}}{T^3} \times \nonumber \\
	& &\biggl\{eB(\rho_\mathrm{ cm})T\coth{(eB(\rho_\mathrm{ cm})T)} \\
        & & 
	- 1 -\frac{1}{3}(eB(\rho_\mathrm{ cm})T)^2\biggr\}. \nonumber
\ea

\begin{figure}
	\centering
		\includegraphics[width=\linewidth,clip,trim=0.8cm 0.3cm 1.2cm 1cm]{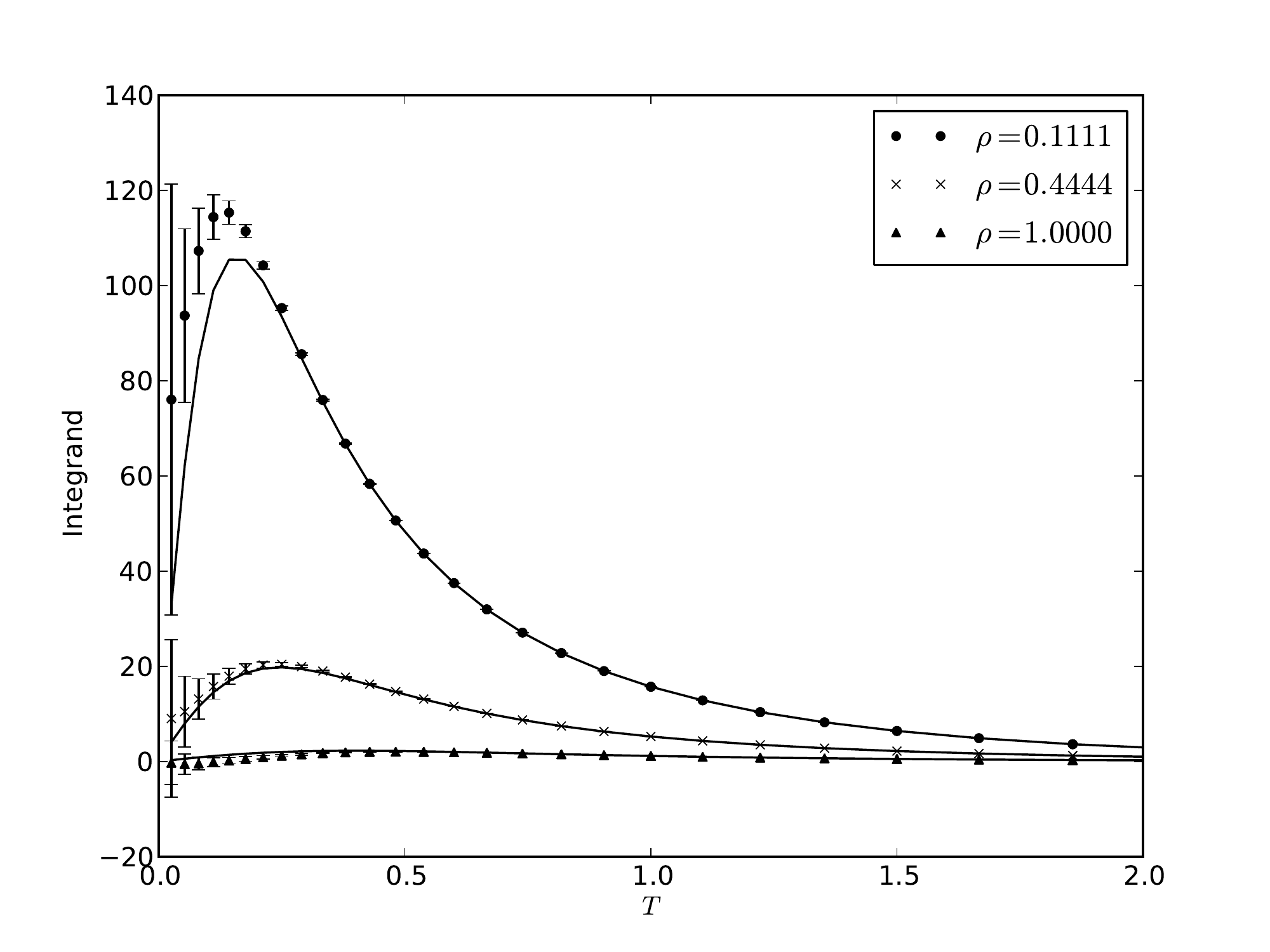}
	\caption[Comparison with derivative expansion for $T$ integrand]
	{The integrand of the proper time, $T$, integral for three different values 
	of the radial coordinate, $\rho$ for a $\lambda = 1$ flux tube. The solid lines 
	represent the zeroth-order derivative expansion, which, as expected, is a good approximation 
	until $\rho$ becomes too small.}
	\label{fig:igrandvsT}
\end{figure}

Figure \ref{fig:igrandvsT} shows a comparison between the proper time integrand,
\be
	\frac{e^{-m^2T}}{T^3}\left[\langle W\rangle_{\vec{r}_\mathrm{ cm}} 
		- 1 -\frac{1}{3}(eB_\mathrm{ cm}T)^2\right],
\ee
and the \ac{LCF} approximation result for a flux tube with $\lambda = \lambda_e$ and
$\mathcal{F} = 10$. In this case, the \ac{LCF} approximation is only appropriate far from the 
center of the flux tube, where the field is not changing very rapidly. In the figure, we can 
begin to see the deviation from this approximation, which gets more pronounced closer to the 
center of the flux tube (smaller values of $\rho$).
\begin{figure}
	\centering
		\includegraphics[width=\linewidth,clip,trim=0.2cm 0.3cm 1.8cm 1cm]{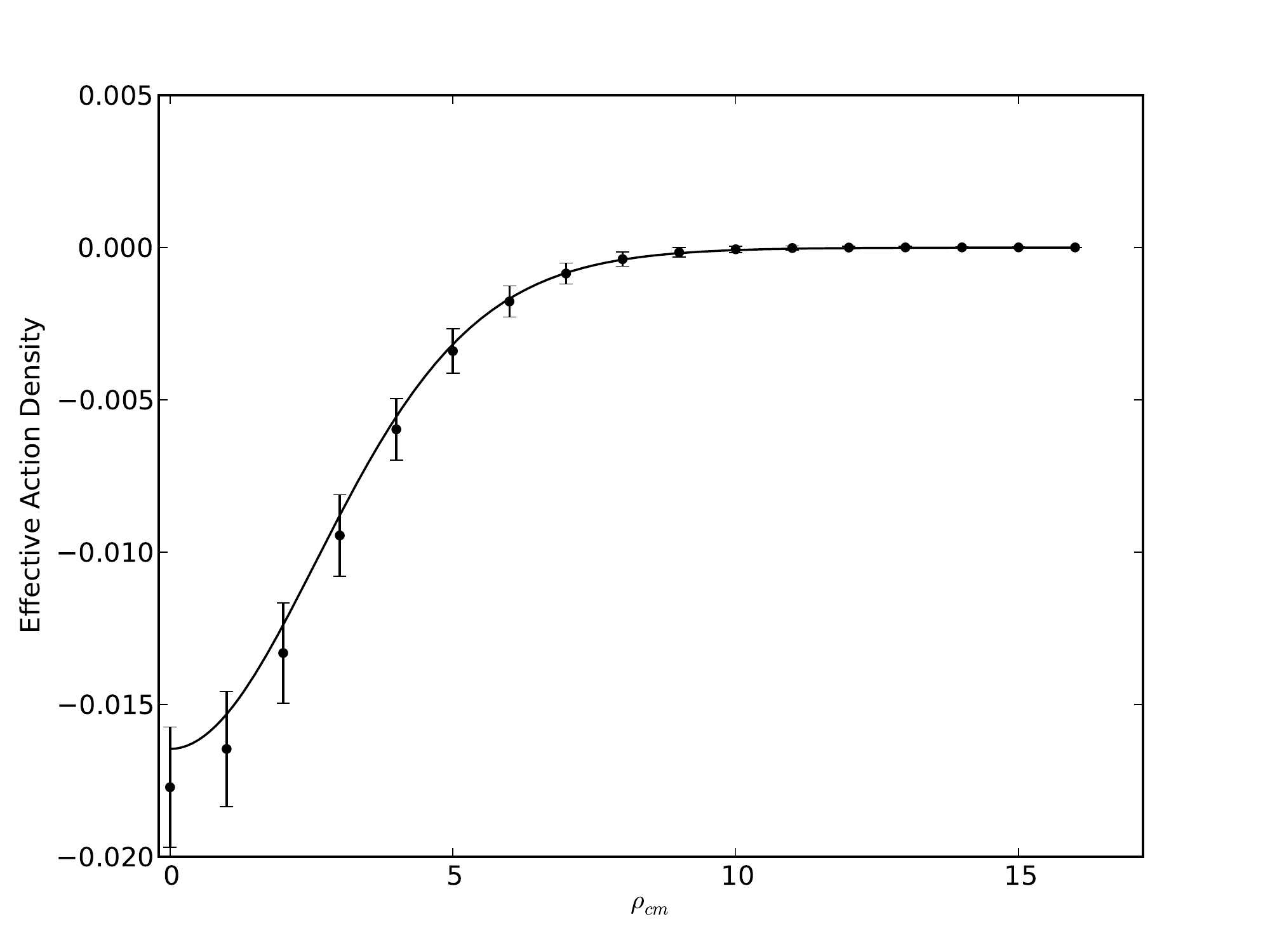}
	\caption[Comparison with \acs{LCF} approximation for action density]
	{The fermion term of the effective action density as a function 
	of the radial coordinate for a flux tube with width $\lambda = 10 \lambda_e$. }
	\label{fig:vsrhocm}
\end{figure}

The effective action density for a slowly varying flux tube is plotted in figure
\ref{fig:vsrhocm} along with the \ac{LCF} approximation. In this case, 
the \ac{LCF} approximation agrees within the statistics of the worldline numerics.




\section{Astrophysical Background}
\label{sec:background}

\subsection{Nuclear Superconductivity in Neutron Stars}
\label{subsec:superconductivity}
In the dense nuclear matter of a neutron star, it may be possible to
have neutron superfluidity, proton superconductivity, and even quark
colour superconductivity~\cite{2006pfsb.book..135S, schmitt2010dense}.
The first prediction of neutron-star superfluidity dates back to
Migdal in 1959~\cite{1959NucPh..13..655M}.  The arguments which make
superfluidity seem likely are based on the temperatures of neutron
stars. A short time after their creation, neutron stars are very cold
compared to nuclear energy scales. The temperature in the interior may
be a few hundred keV. Studies of nuclear matter show that the
transition temperature is $T_c \gtrsim 500$
keV~\cite{1989tns..conf..457S}.  So, it is expected that the nuclear
matter in a neutron star forms condensates of Cooper pairs.

Because of this, some fraction of neutrons in the inner crust 
of a neutron star are expected to be superfluid. These neutrons make up 
about a percent of the moment of inertia of the star and are 
weakly coupled to the nuclear crystal lattice which makes up the 
remainder of the inner crust. If the neutron vortices in the 
superfluid component move at nearly the same speed as the nuclear 
lattice, the vortices can become pinned to the lattice so that the 
superfluid shares an angular velocity with the crust. 
This pinning between the fluid and solid crust has observable impacts 
on the rotational dynamics of the neutron star, for the same reasons that 
hard-boiling an egg (pinning the yolk to the shell) produces an 
observable difference in the way it spins.

This picture of a neutron superfluid co-rotating with a solid crust 
has been used to interpret several types of pulsar timing anomalies.
Pulsars are nearly perfect clocks, although they gradually spin-down 
as they radiate energy. Occasionally, though, pulsars demonstrate 
deviations from their expected regularity. A glitch is an abrupt 
increase in the rotation and spin-down rate of a pulsar, followed 
by a slow relaxation to pre-glitch values over weeks or years. This 
behaviour is consistent with the neutron superfluid suddenly 
becoming unpinned from the crust and then dynamically relaxing 
due to its weak coupling to the crust until it is pinned once 
again~\cite{1975Natur.256...25A}.

The neutron star in Cassiopeia A has been observed to be rapidly cooling
\cite{2010ApJ...719L.167H}. The surface temperature has decreased by about 
4\% over 10 years. This observation is also strong evidence 
of superfluidity and superconductivity in neutron stars
\cite{2011PhRvL.106h1101P, 2011MNRAS.412L.108S}. The observed cooling 
is too fast to be explained by the observed x-ray emissions 
and standard neutrino cooling. However, 
the cooling is readily explained by the emission of neutrinos during 
the formation of neutron Cooper pairs. Based on such a model, 
the superfluid transition temperature of neutron star matter is
$\sim 10^{9}$ K or $\sim 90$ keV. 

Further hints regarding superfluidity in neutron stars come from
long-term periodic variability in pulsar timing data. For example,
variabilities in PSR B1828-11 were initially interpreted as free
precession (or wobble) of the star~\cite{2000Natur.406..484S}.  If
neutron stars can precess, observations could strongly constrain the
ratio of the moments of inertia of the crust and the superfluid
neutrons. Moreover, the existence of flux tubes (\ie type-II
superconductivity) in the crust is generally incompatible with the
slow, large amplitude precession suggested by PSR
B1828-11~\cite{2000Natur.406..484S}. The neutron vortices would have
to pass through the flux tubes, which should cause a huge dissipation
of energy and a dampening of the precession which is not
observed~\cite{PhysRevLett.91.101101}.  However, recent arguments
suggest that the timing variability data is not well explained by free
precession and that it more likely suggests that the star is switching
between two magnetospheric states~\cite{2010Sci...329..408L}.
Nevertheless, other authors suggest not being premature in throwing
out the precession hypothesis without further observations
\cite{2012MNRAS.420.2325J}.

\subsection{Magnetic Flux Tubes in Neutron Stars}

If a magnetic field is able to penetrate the proton superfluid on a
microscopic level, it must do so by forming a triangular Abrikosov
lattice with a single quanta of flux in each flux tube. So, the
density of flux tubes is given simply by the average field strength.
If the distance between flux tubes is $l_f$, the flux in a circular
region within $l_f/2$ of a flux tube is given by
\be
	F = \frac{2\pi \mathcal{F}}{e} = 2\pi \int_0^{l_f/2} B \rho d\rho
\ee
where we have introduced a dimensionless measure of flux $\mathcal{F}
= \frac{e}{2\pi}F$.  So, the distance between flux tubes is
\be
	l_f = \sqrt{\frac{8 \mathcal{F}}{eB}}.
\ee
If the magnetic field is the quantum critical field strength, 
$B_k = \frac{m^2}{e} = 4.4 \times 10^{13}$ Gauss, then 
the flux tubes are separated by a few Compton electron wavelengths.
This is particularly interesting since this is the distance scale 
associated with non-locality in \ac{QED}.

The size of a flux tube profile in laboratory superconductors is on
the nanometer or micron
scale~\cite{poole2007superconductivity}. Because the flux is fixed,
the size of the tube profile determines the strength of the magnetic
field within the tube. For laboratory superconductors the field
strengths are small compared to the quantum critical field, and the
field is slowly varying on the scale of the Compton wavelength.  In
this case, the quantum corrections to the free energy are known to be
much smaller than the classical contribution (see section
\ref{sec:EAisoflux}).  The size scale for the flux tubes in a
superconductor is determined by the London penetration depth.  In a
neutron star, this quantity is estimated to be a small fraction of a
Compton wavelength, much smaller than in laboratory superconductors
\cite{lrr-2008-10, PhysRevLett.91.101101}. In this case, the magnetic
field strength at the centre of the tube exceeds the quantum critical
field strength and the field varies rapidly, rendering the derivative
expansion description of the effective action unreliable.

The Ginzburg-Landau parameter, equation is the 
ratio of the proton coherence length, $\xi_p \sim 30$ fm, and the 
London penetration depth of a proton superconductor, $\lambda_p \sim 80$~fm
\cite{PhysRevLett.91.101101}. 
\be
	\kappa = \frac{\lambda_p}{\xi_p} \sim 2
\ee
where $\kappa>1/\sqrt{2}$ signals Type-II behavior.        
We therefore expect that the proton Cooper pairs most likely form a
type-II superconductor~\cite{2006pfsb.book..135S}.  However, it is
possible that physics beyond what is taken into account in the
standard picture affects the free energy of a magnetic flux tube. In
that case, the interaction between two flux tubes may indeed be
attractive in which case the neutron star would be a type-I
superconductor.

\subsection{\texorpdfstring{\acs{QED}}{QED} Effective Actions of Flux Tubes}
\label{sec:EAisoflux}

Vortices of magnetic flux have very important impacts on the quantum
mechanics of electrons. In particular, the phase of the electron's
wavefunction is not unique in such a magnetic field. This is
demonstrated by the Aharonov-Bohm
effect~\cite{0370-1301-62-1-303,PhysRev.115.485}. The first
calculations of the fermion effective energies of these configurations
were for infinitely thin Aharonov-Bohm flux
strings~\cite{Gornicki1990271, 1998MPLA...13..379S}. Calculations for
thin strings were also performed for cosmic string
configurations~\cite{0264-9381-12-5-013}. For these infinitely-thin
string magnetic fields, the energy density is singular for small
radii. So, it is not possible to define a total energy per unit
length. Another approach was to compute the effective action for a
finite-radius flux tube where the magnetic flux was confined entirely
to the surface of the tube~\cite{PhysRevD.51.810}. This approach
results in infinite classical energy densities as well.

Physical flux tube configurations would have a finite radius.  The
earliest paper to deal with finite radius flux tubes in QED considered
the effective action of a step-function profiled flux tube using the
Jost function of the related scattering
problem~\cite{1999PhRvD..60j5019B}. One of the conclusions from this
research was that the quantum correction to the classical energy was
relatively small for any value of the flux tube size, for the entire
range of applicability of \ac{QED}.  The techniques from this study
were soon generalized to other field profiles including, a
delta-function cylindrical shell magnetic
field~\cite{PhysRevD.62.085024}, and more realistic flux tube
configurations such as the Gaussian~\cite{2001PhRvD..64j5011P} and the
Nielsen-Olesen vortex~\cite{2003PhRvD..68f5026B}.  Flux tube vacuum
energies were also analyzed extensively using a spectral method
\cite{2005NuPhB.707..233G, 2006JPhA...39.6799W, Weigel:2010pf}.

The effective actions of flux tubes have been previously analyzed
using worldline numerics~\cite{Langfeld:2002vy}.  This research
investigated isolated flux tubes, but also made use of the loop cloud
method's applicability to situations of low symmetry to investigate
pairs of interacting vortices. One conclusion from that investigation
was that the fermionic effects resulted in an attractive force between
vortices with parallel orientations, and a repulsive force between
vortices with anti-parallel orientations.  Due to the similarity in
scope and technique, the latter mentioned research is the closest to
the research presented in this paper.

\section{Calculations}
\label{sec:calculations}

In this section, we will further explore the nature of this phenomenon
in \ac{QED} using a highly parallel implementation of the worldline numerics
technique implemented on a heterogeneous \ac{CPU} and \ac{GPU} architecture. 
Specifically, we
explore cylindrically symmetric magnetic field profiles for isolated
flux tubes and periodic profiles designed to model properties of a
triangular lattice. For these calculations, the classical magnetic field 
configurations are a chosen input to the algorithms and the physical processes that 
may have created the field configurations do not factor in to the calculations. 
The worldline numerics algorithm cannot be straight-forwardly applied to spinor
\ac{QED} calculations in our model lattice because of the well-known
fermion problem of worldline numerics (see section \ref{sec:fermionproblem})
~\cite{Gies:2001zp, MoyaertsLaurent:2004}. 
However, the problem does not affect the scalar QED (\ac{ScQED})
 calculations. Therefore, we explore the quantum-corrected
energies of isolated flux tubes for both scalar and spinor electrons
and use this comparison to speculate about the relationship of our
cylindrical lattice model and the spinor \ac{QED} energies of an
Abrikosov lattice of flux tubes that may be found in neutron stars.



\subsection{Isolated Flux Tubes}

As discussed in section~\ref{sec:cylindrical} we will focus on fields
with a cylindrical symmetry.  In particular to explore isolated
magnetic flux tubes, we consider the following profile function (as we
used earlier in section~\ref{sec:verify}):
\be
f_\lambda(\rho^2) =
\frac{\rho^2}{\rho^2 + \lambda^2}.
\ee
This gives a magnetic field with a profile
\be
\label{eqn:isoB}
B_z(\rho^2) = \frac{2\mathcal{F}}{e}\frac{\lambda^2}{(\rho^2+\lambda^2)^2}.
\ee
This profile is a smooth flux tube representation that can be evaluated quickly.
Moreover, flux tubes with this profile were studied previously using worldline numerics
\cite{Moyaerts:2003ts, MoyaertsLaurent:2004}.

\subsection{Cylindrical Model of a Flux Tube Lattice}

\begin{figure}
	\centering
		\includegraphics[width=\linewidth]{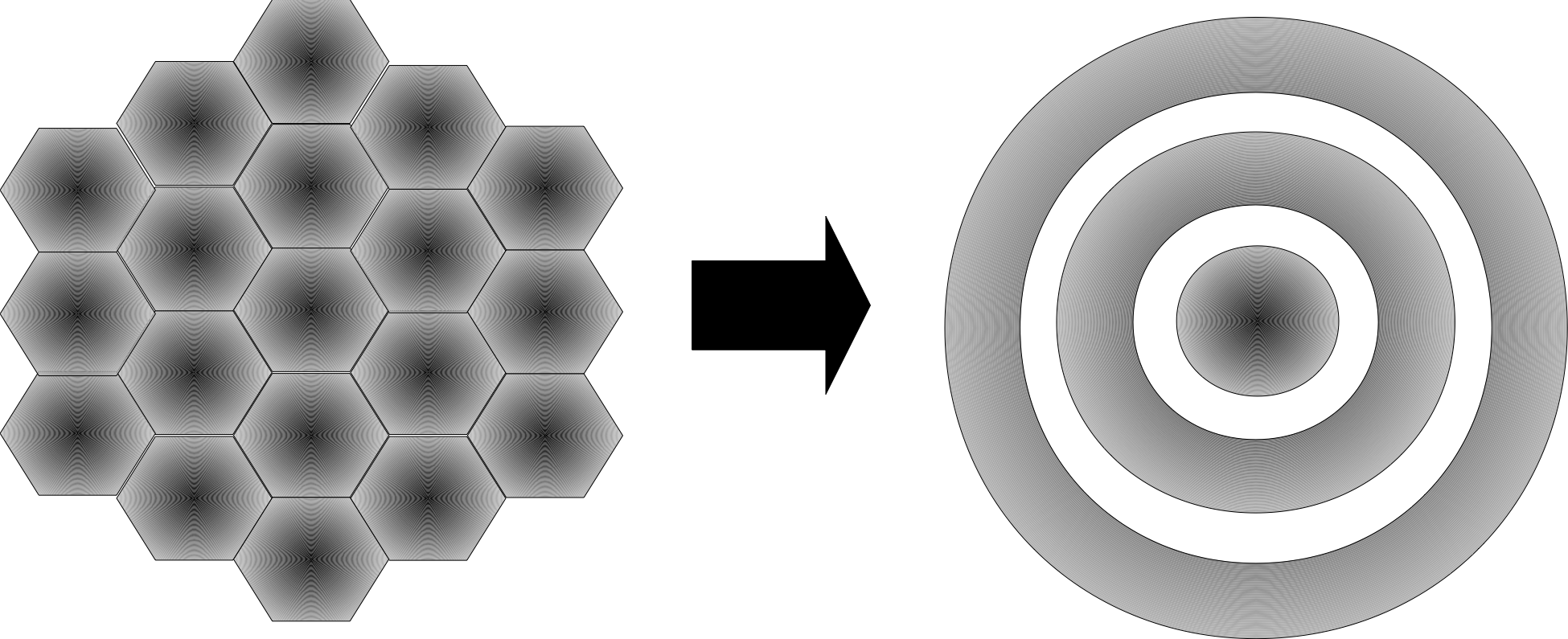}
	\caption[Cylindrical model of a hexagonal lattice]
	{In a type-II superconductor, there are neighbouring flux 
	tubes arranged in a hexagonal Abrikosov lattice which have a non-local 
	impact on the effective action of the central flux tube (left). In our model, 
	we account for the contributions from these neighbouring flux tubes in 
	cylindrical symmetry by including surrounding rings of flux (right).}
	\label{fig:latticesymmetry}
\end{figure}

In a neutron star, we do not have isolated flux tubes. The tubes are
likely arranged in a dense lattice with the spacing between tubes on
the order of the Compton wavelength, with the size of a flux tube a
few percent of the Compton wavelength. Specifically, the maximum size
of a flux tube is on the order of the coherence length of the
superconductor, which for neutron stars has been estimated to be $\xi
\approx 30$ fm~\cite{PhysRevLett.91.101101}.  This situation can be
directly computed in the worldline numerics technique. However, this requires us
to integrate over two spatial dimensions instead of one. Moreover, it
requires the use of more loops to more precisely probe the spatial
configurations of the magnetic field. Despite these problems, it is
very interesting to consider a dense flux tube lattice. Unlike the
isolated flux tube, the wide-tube limit of the configuration doesn't
have zero field, but an average, uniform background field. If this
background field is the size of the critical field, there are
interesting quantum effects even in the wide-tube limit.

In this section, we build a cylindrically symmetric toy model of a
hexagonal flux tube lattice. We focus on one central flux tube and
treat the surrounding six flux tubes as a continuous ring with six
units of flux at a distance $a$ from the central tube. The next ring
will contain twelve units of flux at a distance of $2a$, etc (see
figure \ref{fig:latticesymmetry}).  Because of this condition, the
average strength of the field is fixed, and the field becomes uniform
in the wide tube limit instead of going to zero. For small values of
$\lambda$, we will have non-local contributions from the surrounding
rings in addition to the local contributions from the central flux
tube. This strategy will result in a simple model relative to the 
sophisticated flux tube lattice models used in the 
context of superconducting physics. However, the simplifications 
are appropriate for a first study of the non-local QED interactions
between the regions of flux.

It is difficult to construct a model of this scenario if the flux
tubes bleed into one another as they are placed close together. For
example, with Gaussian flux tubes or flux tubes with the profile used
in the previous section, it is difficult to increase the width of the
flux tubes while accounting for the magnetic flux that bleeds out of
their regions. Moreover, it is difficult to integrate these schemes to
find the profile function $f_\lambda(\rho)$ which is needed to compute
the scalar part of the Wilson loop.  In order to keep each tube as a
distinct entity which stays within its assigned region, we assign a
smooth function with compact support to represent each tube.  This is
most easily done with the bump function, $\Psi(x)$, defined as
\be
	\Psi(x) = 
	\begin{cases}
	e^{-1/(1-x^2)} & \mbox{ for } |x| < 1\\
	0 & \mbox{ otherwise} 
	\end{cases}.
\ee
This function can be viewed as a rescaled Gaussian.

We start by defining the magnetic field outside of the central flux
tube. Here, the magnetic field is a constant background field, with
the flux ring contributing a bump of width $\lambda$. The height of
the bump must go to zero as the width of the flux tube approaches the
distance between flux tubes, and should become infinite as the flux
tube width goes to zero:
\be
	B_z(\rho>\frac{a}{2}) = B_{\rm bg} + A\left(\frac{a-\lambda}{\lambda}\right)\left[\Psi(2(\rho-n a)/\lambda)-B\right].
\ee
with $n\equiv \lfloor\frac{\rho+a/2}{a}\rfloor$.

If we require 6 units of flux in the first outer ring, 12 in the second, 
and so on (see figure \ref{fig:latticesymmetry}), the size of the background field 
is fixed to $B_{\rm bg} = \frac{6 \mathcal{F}}{ea^2}$. The total flux contribution due to the $\lambda$-dependent 
terms must be zero:

\be
	\int_{(n-1/2)a}^{(n+1/2)a} \rho A\left(\frac{a-\lambda}{\lambda}\right)
		\left[\Psi(2(\rho-n a)/\lambda)-B\right] d\rho = 0
\ee

\be
	\frac{\lambda}{2}\int_{-1}^1 \left(\frac{\lambda}{2}x+na\right)\Psi(x)dx-Ba^2n = 0.
\ee
This fixes the value of the constant $B$ to
\be
	B=\frac{q_1}{2}\frac{\lambda}{a}.
\ee
The numerical constant $q_1$ is defined by
\be
	q_1 = \int_{-1}^{1}\Psi(x)dx \approx 0.443991.
\ee
For a given bump amplitude, $A$, the magnetic field will become
negative if $\lambda$ becomes small enough.  Therefore, we replace $A$
with its maximum value for which the field is positive if $\lambda >
\lambda_{\rm min}$ for some choice of minimum flux tube size:
\be
	A=\frac{12 \mathcal{F}}{e a q_1 (a-\lambda_{\rm min})}.
\ee
The choice of $\lambda_{\rm min}$ sets the tube width at which the
field between the flux tubes vanishes. If $\lambda < \lambda_{\rm
  min}$, the magnetic field between the flux tubes will point in the
$-\hat{\vec{z}}$-direction.

\begin{figure}
	\centering
		\includegraphics[width=\linewidth,clip,trim=0.45in 0.1in 0.55in 0.3in]{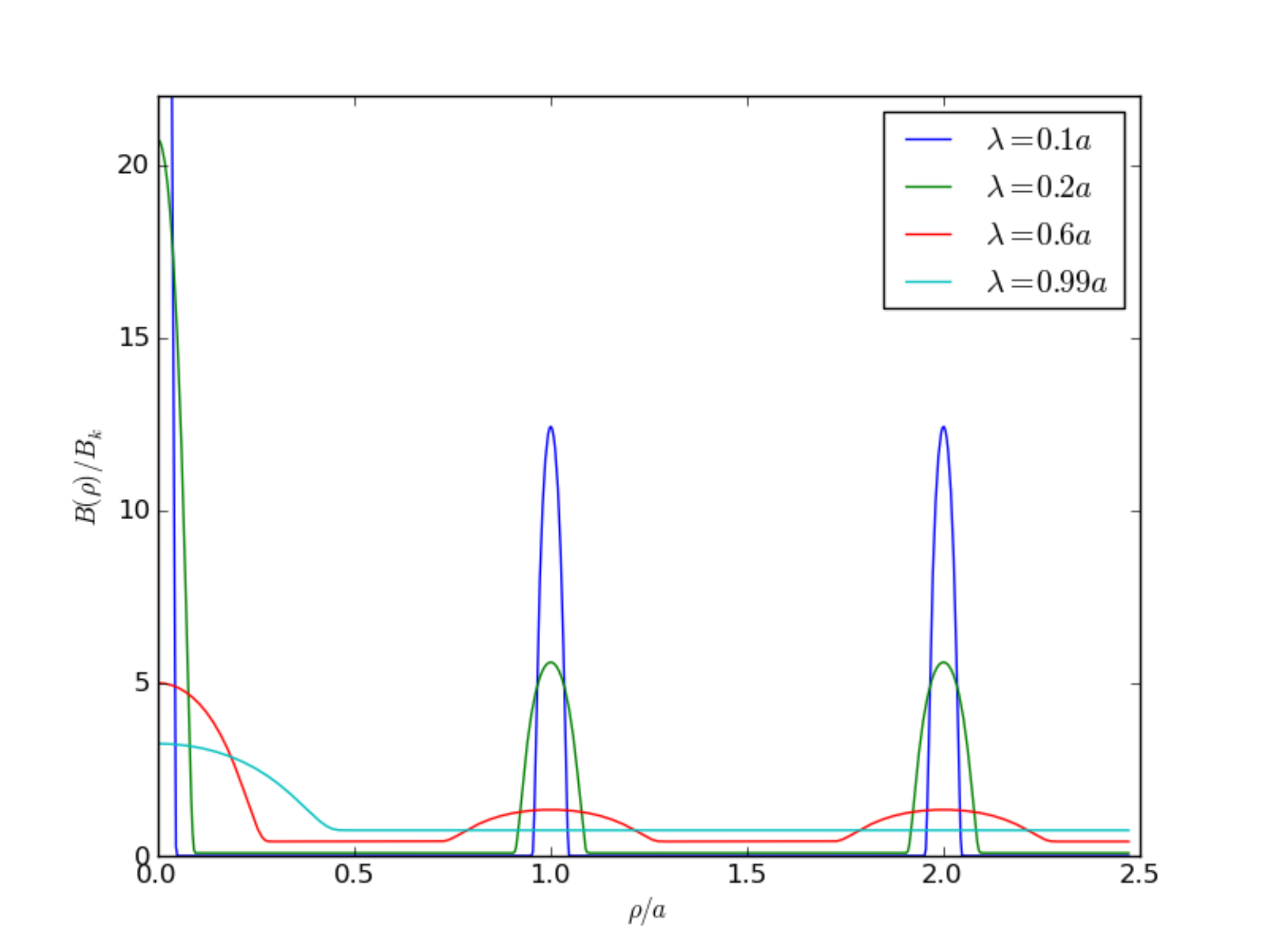}
	\caption[Magnetic field in a cylindrical lattice model]
	{The cylindrical lattice flux tube model for several 
	values of the width parameter $\lambda$. Here we have taken $a =\sqrt{8}\lambda_e$
	and $\lambda_{\rm min} = 0.1a$. Note that in the limit $\lambda\rightarrow a$
	the field is nearly uniform with a mound in the central region. This is a 
	consequence of the flux conditions in cylindrical symmetry requiring different
	fields in the internal and external regions. The height of the $\lambda=0.1a$
	flux tube extends beyond the height of the graph to about $61.9B_k$. 
	A 3D surface plot of the $\lambda=0.6a$ field profile is shown in figure 
	\ref{fig:B3d}.}
	\label{fig:periodicB}
\end{figure}

\begin{figure}
	\centering
		\includegraphics[width=\linewidth,clip,trim=0.45in 0.1in 0.55in 0.3in]{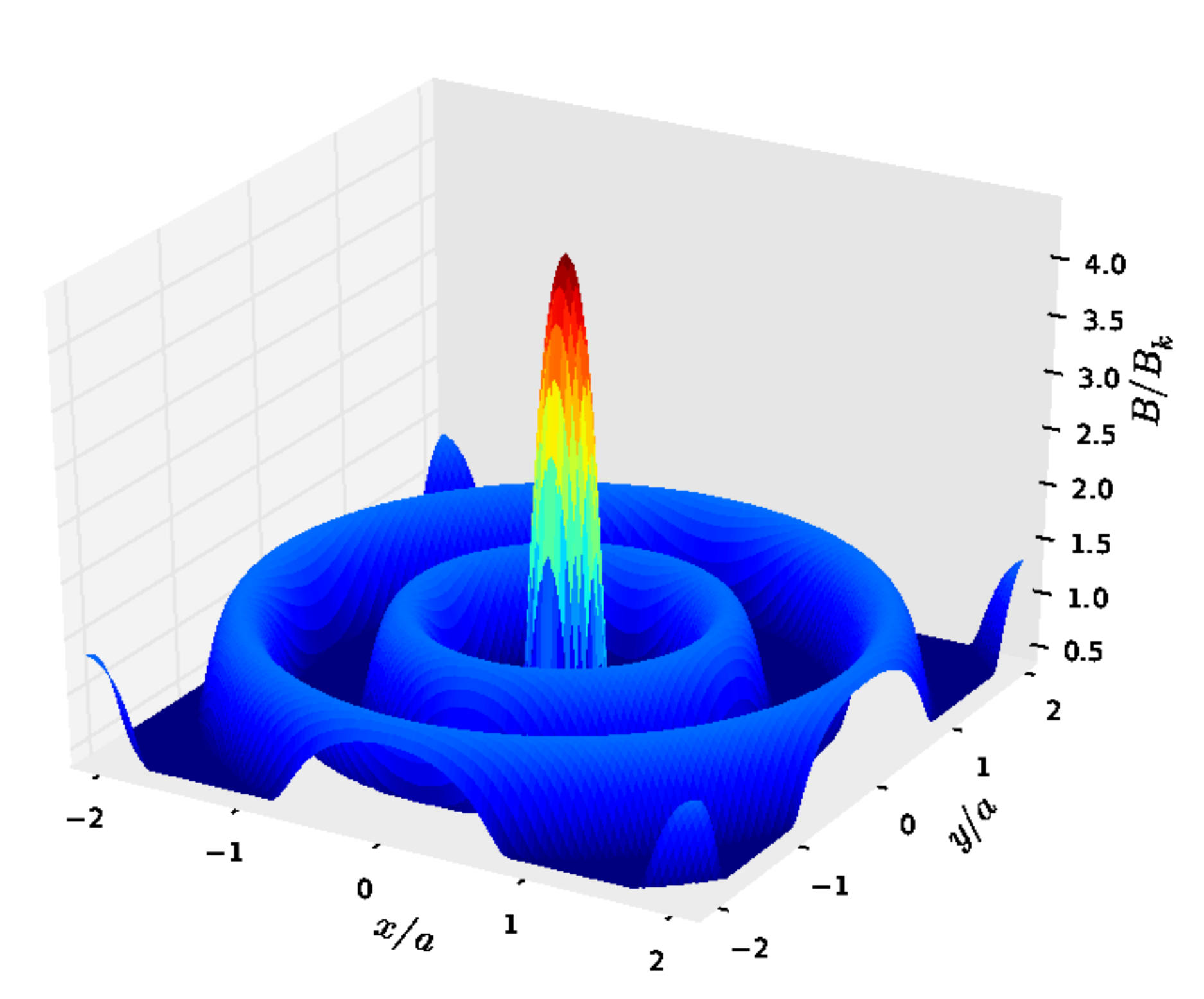}
	\caption[Surface plot of magnetic field model]
	{A 3D surface plot of the $\lambda = 0.6a$ (red in figure \ref{fig:periodicB}) magnetic field profile from 
	figure \ref{fig:periodicB}.}
	\label{fig:B3d}
\end{figure}

\begin{figure}
	\centering
		\includegraphics[width=\linewidth,clip,trim=0.4in 0.1in 0.55in 0.3in]{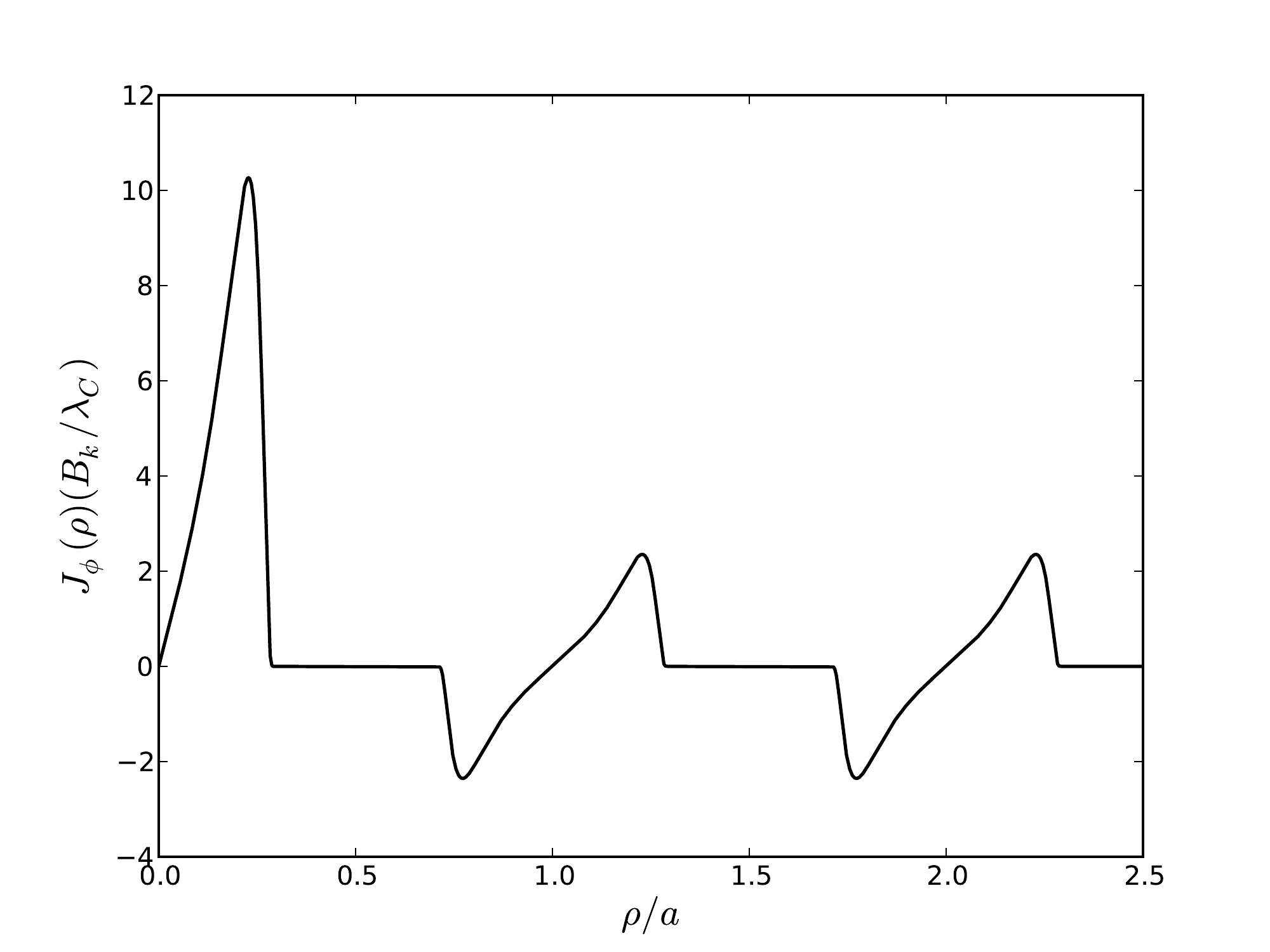}
	\caption[Classical current required to create magnetic field model]
	{The current densities required to create the $\lambda=0.6a$ (red) magnetic field profile. 
	The current is given by the curl of the magnetic field, $J_\psi(\rho) = -\frac{d B_z(\rho)}{d\rho}$.
	The conversion to SI units is $1B_k/\lambda_e \approx 6\times 10^{38}{\rm A/m}^3$.}
	\label{fig:current}
\end{figure}

Because we are trying to fit a hexagonal peg into a round hole, we
must treat the central flux tube differently.  For example, the
average field inside the central region for a unit of flux, is
different than the average field in the exterior region. Therefore,
even when $\lambda \rightarrow a$, the field cannot be quite uniform.
We consider the field in the central region to be a constant field
with a bump centered at $\rho = 0$:
\be
	B_z(\rho < \frac{a}{2}) = A_0 \Psi(2\rho/\lambda) + B_0.
\ee
The constant term, $B_0$ is fixed by requiring continuity with the
exterior field at $\rho = a/2$:
\be
	B_0 = \frac{6\mathcal{F}}{e a^2}\left(1-\frac{a-\lambda}{a-\lambda_{\rm min}}\right).
\ee
The bump amplitude, $A_0$, is determined by fixing the flux in the central region,
\be
	\int_0^{a/2}\rho\left[A_0\Psi(2\rho/\lambda) + B_0\right]d\rho = \frac{\mathcal{F}}{e}:
\ee
\be
	A_0 \left(\frac{\lambda}{2}\right)^2\int_0^1 x \Psi(x)dx + 
		\frac{B_0}{2}\left(\frac{a}{2}\right)^2 = \frac{\mathcal{F}}{e}
\ee
\be
	A_0 = \frac{4 \mathcal{F}}{\lambda^2 e q_2}\left(1-\frac{3}{4}
		\left(1-\frac{a-\lambda}{a-\lambda_{\rm min}}\right)\right),
\ee
where the numerical constant, $q_2$, is defined by
\be
	\label{eqn:q2}
	q_2 \equiv \int_0^1 x \Psi(x) dx \approx 0.0742478.
\ee
Finally, collecting together the important expressions, 
the cylindrically symmetric flux tube lattice model is
\ba
\label{eqn:ffbless}
B_z(\rho\le\frac{a}{2}) &=& \frac{4 \mathcal{F}}{\lambda^2 e q_2}
	\left(1-\frac{3}{4}\left(\frac{\lambda-\lambda_{\rm min}}{a-\lambda_{\rm min}}\right)\right)\Psi(2\rho/\lambda) \nonumber \\
	& & + \frac{6\mathcal{F}}{ea^2}\left(\frac{\lambda-\lambda_{\rm min}}{a-\lambda_{\rm min}}\right)
\label{eqn:Bzext} \\
B_z(\rho>\frac{a}{2}) &=& \frac{6 \mathcal{F}}{e a^2}\left(\frac{\lambda-\lambda_{\rm min}}{a-\lambda_{\rm min}}\right) + \nonumber \\
	& & \frac{12 \mathcal{F}}{q_1ea\lambda}\left(\frac{a-\lambda}{a-\lambda_{\min}}\right)
	\Psi(2(\rho-na)/\lambda)  .
\ea
The magnetic field profile defined by these equations is shown in figures
\ref{fig:periodicB} and \ref{fig:B3d}. The current density required to 
created fields with this profile is shown in figure \ref{fig:current}.

\subsection{The Classical Action}

The classical action, $\Gamma^0$, is infinite for this configuration. To obtain 
a finite result, we must look at the action per unit length in the z-direction, 
per unit time, and per flux tube region (\ie for $\rho < a/2$). 
The action for such a region in 
cylindrical coordinates is given by 

\be
	\frac{\Gamma^0}{T L_z} = -\pi \int_0^{a/2}\rho B_z(\rho)^2d\rho.
\ee
After substituting the value of equation (\ref{eqn:ffbless}), the magnetic field in the interior 
region:
\ba
	\frac{\Gamma^0}{T L_z} &=& -\pi \int_0^{a/2}\rho \biggr[
	\frac{4\mathcal{F}}{\lambda^2e^2q_2}\left(1-\frac{3}{4}
	\left(\frac{\lambda-\lambda_{\rm min}}{a-\lambda_{\rm min}}\right)\right)
	\Psi(2\rho/\lambda)\nonumber \\
	& & + \frac{6\mathcal{F}}{ea^2}\left(\frac{\lambda-\lambda_{\rm min}}{a-\lambda_{\rm min}}\right)
	\biggr]^2 d\rho.
\ea
After some algebra, we are left with an expression for the classical action,
\ba
	\frac{\Gamma_0}{TL_z} &=& \pi \int_0^{a/2}\rho B_z(\rho)^2d\rho \\
	& = & -\frac{\pi \mathcal{F}^2}{e^2a^2}\biggr\{4\frac{a^2q_3}{\lambda^2q_2^2}
	+\left(\frac{\lambda-\lambda_{\rm min}}{a - \lambda_{\rm lmin}}\right)
	\biggr[ \nonumber \\
	& &\left(\frac{9}{4}\frac{a^2q_3}{\lambda^2q_2^2}-\frac{9}{2}\right)
	\left(\frac{\lambda-\lambda_{\rm min}}{a - \lambda_{\rm lmin}}\right) 
	-6\frac{a^2q_3}{\lambda^2q_2^2}+12 \biggr]\biggr\},
\ea
where $q_3$ is another numerical constant related to integrating the bump function:
\be
	\label{eqn:q3}
	q_3 \equiv \int_0^1 x\left(\Psi(x)\right)^2 dx 
	= \int_0^1 x e^{-\frac{2}{1-x^2}}dx \approx 0.0187671.
\ee

\subsection{Integrating to Find the Potential Function}

To compute the Wilson loops, it is generally required to use the
vector potential which describes the magnetic field. For us, this
means that we must find $f_\lambda(\rho)$ for our magnetic field
model.  This could always be done numerically, but can be
computationally costly since it is evaluated by every discrete point
of every worldline in the ensemble.  For computations on the \ac{CUDA}
device, an increase in the complexity of the kernel often means that
less memory resources are available per processing thread, limiting
the number of threads that can be computed concurrently.  It is
therefore preferable to find an analytic expression for this function.
From equation (\ref{eqn:isoB}), this function is related to the
integral of the magnetic field with respect to $\rho^2$. For the inner
region, we have
\ba
	f_\lambda(\rho < a/2) &=& \frac{e}{2\mathcal{F}}\biggr[\frac{4\mathcal{F}}{\lambda^2eq_2}
		\left(1-\frac{3}{4}\left(\frac{\lambda-\lambda_{\rm min}}{a-\lambda_{\rm min}}\right)\right) \times \nonumber \\
		& & \int_0^{\rho^2}\Psi\left(\frac{2\rho'}{\lambda}\right)d \rho'^2 +  \\
		& &\frac{6 \mathcal{F}}{ea^2}\left(\frac{\lambda-\lambda_{\rm min}}{a-\lambda_{\rm min}}\right)\rho^2\biggr]. \nonumber
\ea
The integral over the bump function can be computed in terms of the exponential integral 
$E_i(x) = \int_{-\infty}^x \frac{e^t}{t} dt$:
\ba
	\int_0^{\rho^2}\Psi\left(\frac{2\rho'}{\lambda}\right)d\rho'^2 &=& 
	\left(\frac{\lambda}{2}\right)^2\Biggr[2q_2+\left(\frac{4\rho^2}{\lambda^2}-1\right) 
		e^{-\frac{1}{1-\frac{4\rho^2}{\lambda^2}}}- \nonumber \\
                & & ~~~ E_i\left(-\frac{1}{1-\frac{4\rho^2}{\lambda^2}}\right)\Biggr]
\ea
for $\rho < \lambda/2$ and
\be
	\int_0^{\rho^2}\Psi\left(\frac{2\rho'}{\lambda}\right)d\rho'^2 = \frac{q_2\lambda}{2}
\ee
for $\rho \ge \lambda/2$. Our expression for the profile function in the inner region is
\be
	f_\lambda(\rho\le a/2) = \left(1-\frac{3}{4}\left(\frac{\lambda-\lambda_{\rm min}}{a-\lambda_{\rm min}}\right)\right)
		\Phi(2\rho/\lambda) + \frac{3\rho^2}{a^2}\left(\frac{\lambda-\lambda_{\rm min}}{a-\lambda_{\rm min}}\right),
\ee
with
\be
	\Phi(x) \equiv
	\begin{cases}
	1 + \frac{1}{2q_2}\left(x^2-1\right)e^{-\frac{1}{1-x^2}}
		-\frac{1}{2q_2}E_i\left(-\frac{1}{1-x^2}\right)& \mbox{ for } x < 1\\
	1 & \mbox{ for } x \ge 1 
	\end{cases}.
\ee

\begin{figure}
	\centering
		\includegraphics[width=\linewidth,clip,trim=0.4in 0.1in 0.55in 0.3in]{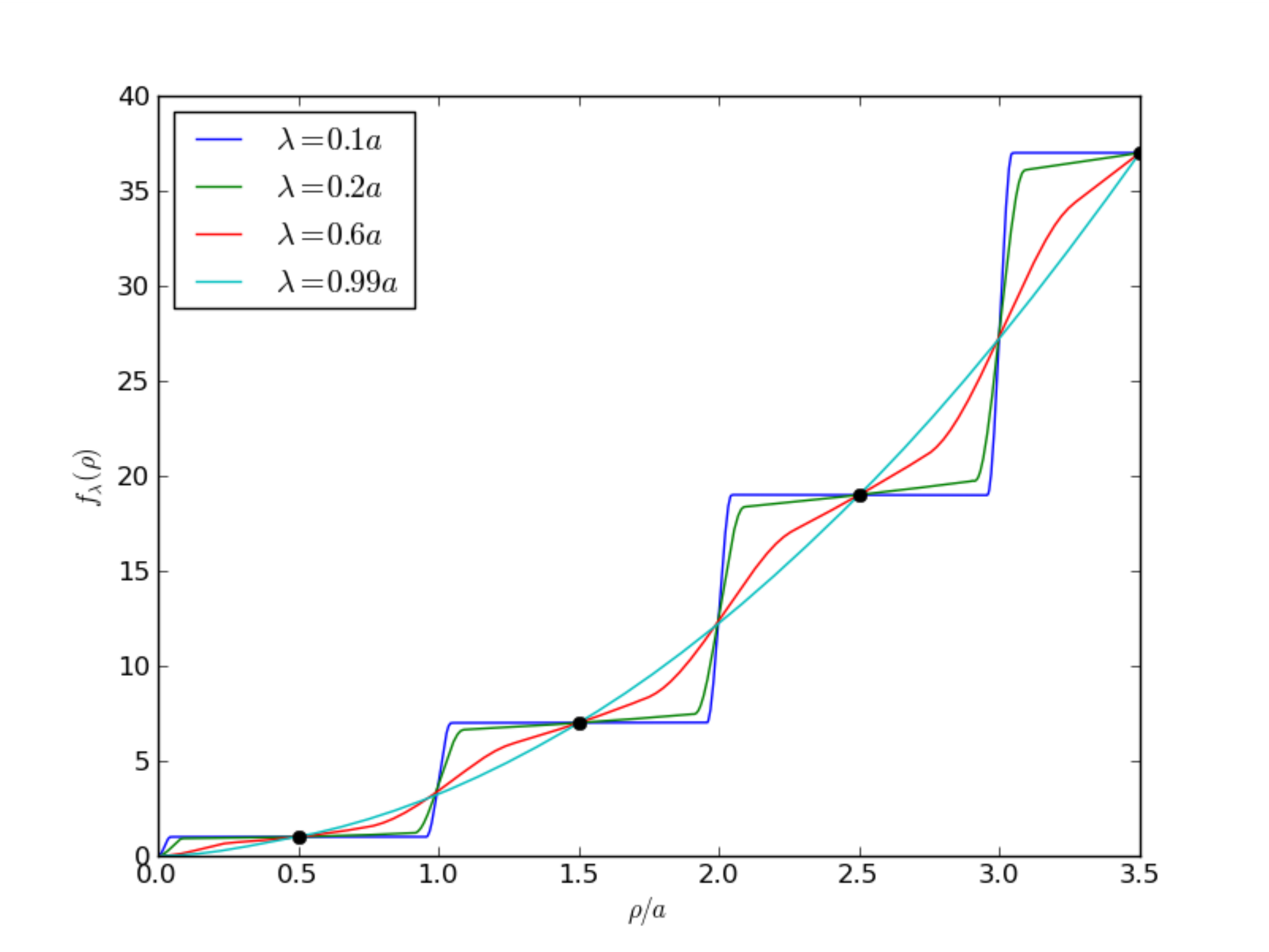}
	\caption[The profile function in a cylindrical lattice model]
	{The function $f_\lambda(\rho)$ for the above described magnetic 
	field model. The flux conditions require the function to pass through the 
	black dots. A quadratic function corresponds to a uniform field while 
	a staircase function corresponds to delta-function flux tubes. The parameter 
	$\lambda$ smoothly makes the transition between these two extremes.
	Each of these functions corresponds to a magnetic field profile in 
	figure \ref{fig:periodicB}.}
	\label{fig:periodicfl}
\end{figure}

The exterior integral is a bit more challenging, but we can make
significant progress and obtain an approximate expression. The first
term is a constant given by the value of the profile function at $\rho
= a/2$. This value is given by the flux in the central flux tube,
which we have already chosen to be 1,
\be
	f_\lambda(\rho>a/2) = 1 + \frac{e}{\mathcal{F}}\int_{a/2}^{\rho}\rho'B(\rho'>a/2)d\rho'.
\ee
We may put the magnetic field, equation (\ref{eqn:Bzext}), into this expression to get
\ba
	f_\lambda(\rho>a/2) &=& 1 + \frac{3}{4}\left(\frac{4\rho^2}{a^2}-1\right)
		\left(\frac{\lambda-\lambda_{\rm min}}{a-\lambda_{\rm min}}\right) +  \\
		& & \frac{12}{q_1a\lambda}\left(\frac{a-\lambda}{a-\lambda_{\rm min}}\right)
		\int_{a/2}^{\rho}\rho'\Psi\left(\frac{2(\rho-na)}{\lambda}\right)d\rho'. \nonumber
\ea
The remaining integral is over every bump between $\rho'=a/2$ and
$\rho'=\rho$. We express the result as a term which accounts for each
completely integrated bump, and an integral over the partial bump if
$\rho$ is within a bump:
\ba
	f_\lambda(\rho>a/2) &=& 1 + \frac{3}{4}\left(\frac{4\rho^2}{a^2}-1\right)
		\left(\frac{\lambda-\lambda_{\rm min}}{a-\lambda_{\rm min}}\right)+ \nonumber \\
		 & & 3n(n-1)\left(\frac{a-\lambda}{a-\lambda_{\rm min}}\right) +  \\
		& & \frac{3\lambda}{q_1a}\left(\frac{a-\lambda}{a-\lambda_{\rm min}}\right)\chi(2(\rho-na)/\lambda), \nonumber
\ea
where
\be
	\chi(x_0) = 
	\begin{cases}
	0 & \mbox{ for } x_0 \le -1  \\
	\int_{-1}^{x_0}xe^{-\frac{1}{1-x^2}}dx 
			+ \frac{2na}{\lambda}\int_{-1}^{x_0}e^{-\frac{1}{1-x^2}}dx& \mbox{ for } |x_0| < 1\\
	 \frac{2naq_1}{\lambda} & \mbox{ for } x_0 \ge 1 	  
	\end{cases}.
\ee
One of the integrals in $\chi(x_0)$ can be expressed in terms of the exponential integral:
\be
	\int_{-1}^{x_0}xe^{-\frac{1}{1-x^2}}dx =\frac{1}{2}\left[(x_0^2-1)e^{-\frac{1}{1-x_0^2}}
		-E_i\left(-\frac{1}{1-x_0^2}\right)\right].
\ee
The remaining integral cannot be simplified analytically. To use this
integral in our numerical model, it must be computed for each discrete
point on each loop for each $\rho_{\rm cm}$ and $T$ value.  Therefore,
it is worthwhile to consider an approximate expression which models
the integral, and can be computed faster than performing a numerical
integral each time. To find this approximation, we computed the
numerical result at 300 values between $x_0=-1.2$ and $x_0=1.2$. The
data was then input into Eureqa Formulize, a symbolic regression
program which uses genetic algorithms to find analytic representations
of arbitrary data~\cite{formulize}.  A similar technique has been used
to produce approximate analytic solutions of ODEs~\cite{geneticODEs}.
The result is a model of the numerical data points with a maximum
error of 0.0001 on the range $|x_0| < 1.0$:
\be
	\int_{-1}^{x_0} e^{-\frac{1}{1-x^2}}dx \approx 
	\frac{0.444}{1+e^{-3.31x_0 - g(x_0)}}
\ee
where
\be
g(x_0) =  
	\frac{5.25x_0^3 - 3.31x_0^2\sin{(x_0)}\cos{(-0.907x_0^2-1.29x_0^8)}}{\cos{(x_0)}}
\ee
This function evaluates ten times faster than the numerical
integral evaluated at the same level of precision with the \ac{GSL}
Gaussian quadrature library functions and with fewer memory registers.
Using this approximation introduces a systematic uncertainty which is
small compared to that associated with the discretization of the loop
integrals, and considerably smaller than the statistical error bars.
Using these expressions for the integrals, we can express
$f_\lambda(\rho)$ in any region in terms of exponential integrals,
exponential, and trigonometric functions with suitable precision.
Furthermore, for computation we may express the exponential integral
as a continued fraction. The profile function, $f_\lambda(\rho)$, is
plotted in figure \ref{fig:periodicfl}.

\section{Results}
\label{sec:periodic_results}

\subsection{Comparing Scalar and Fermionic Effective Actions}

Because of the fermion problem of worldline numerics \cite{Gies:2001zp,
  MoyaertsLaurent:2004}, the 1-loop effective action for the
cylindrical flux tube lattice model was not computed for the case
of spinor \ac{QED}. Performing this calculation in the spinor case 
would require subtle numerical cancellations between large terms, 
and represents a significant numerical challenge. 
Fortunately, the fermion problem does not affect the scalar
case. So, we will analyze this model for \ac{ScQED}. However, 
in this section we will compare the scalar and fermionic effective
actions for isolated flux tubes to demonstrate that the behaviour of 
both theories is qualitatively and numerically similar.

\begin{figure}
	\centering
		\includegraphics[width=\linewidth,clip,trim=0.30in 0.1in 0.55in 0.3in]{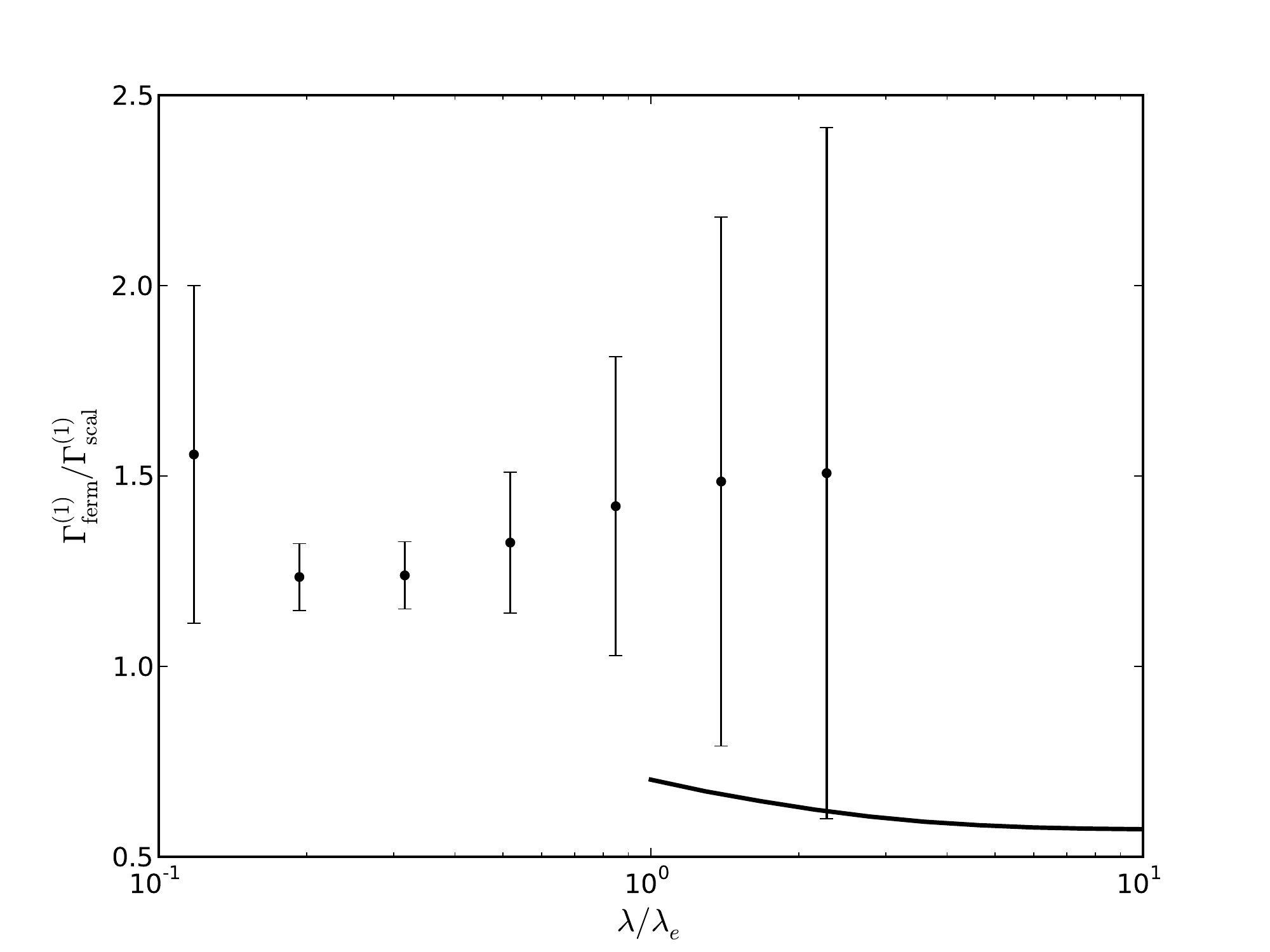}
	\caption[Comparison of 1-loop term in \acs{QED} and \acs{ScQED}]
	{The ratio of the 1-loop term in \ac{QED} to the 1-loop term 
	in \ac{ScQED}. The solid line is the \ac{LCF} approximation, 
	while the points are the result of worldline numerics calculations. Note that the 
	\ac{LCF} approximation breaks down near $\lambda = \lambda_e$ and 
	that the statistics from point to point are strongly correlated. This plot shows that 
	the 1-loop correction in \ac{ScQED} differs from the \ac{QED} correction by a factor 
	close to unity for a wide range of flux tube widths.}
	\label{fig:scalfermiratio}
\end{figure}

For isolated flux tubes, the decay of the magnetic field for large distances
protects the calculations from the fermion problem. Therefore, the
effective action can be computed for both scalar and spinor 
\ac{QED}. In figure \ref{fig:scalfermiratio}, we plot the 
ratio of the spinor to scalar 1-loop correction term for 
identical magnetic fields, along with the prediction of the 
\ac{LCF} approximation for large values of $\lambda$. The 
\ac{LCF} approximation in \ac{ScQED} is given by
\ba
	\label{eqn:LCFscal}
	\Gamma^{(1)}_{\rm scal} &=& -\frac{1}{2\pi}\int_0^\infty dT 
	\int_0^\infty \rho_{\rm cm} d\rho_{\rm cm}\frac{e^{-m^2T}}{T^3} \nonumber \\
	& &\left\{\frac{eB(\rho_{\rm cm})T}{\sinh{(eB(\rho_{\rm cm})T)} }
	- 1 +\frac{1}{6}(eB(\rho_{\rm cm})T)^2\right\}.
\ea
This can be compared to the spinor \ac{QED} approximation, 
\ba
	\label{eqn:LCFferm2}
	\Gamma^{(1)}_\mathrm{ ferm} &=& \frac{1}{4\pi}\int_0^\infty dT 
	\int_0^\infty \rho_\mathrm{ cm} d\rho_\mathrm{ cm}\frac{e^{-m^2T}}{T^3} \times \nonumber \\
	& &\biggl\{eB(\rho_\mathrm{ cm})T\coth{(eB(\rho_\mathrm{ cm})T)} \\
        & & 
	- 1 -\frac{1}{3}(eB(\rho_\mathrm{ cm})T)^2\biggr\}. \nonumber
\ea

There are two important notes to make about figure \ref{fig:scalfermiratio}. 
Firstly, the \ac{LCF} approximation is only a good approximation 
for $\lambda \gg \lambda_e$, and isn't accurate when pushed near its formal 
validity limits~\cite{MoyaertsLaurent:2004}. The second note is that the 
statistics of the points computed with worldline numerics are strongly correlated. So, we conclude 
that the \ac{ScQED} 1-loop correction is larger than the \ac{QED} correction 
for large $\lambda$, and this appears to be reversed for small $\lambda$. 
However, the large worldline numerics error bars and the invalidity of the 
\ac{LCF} approximation near $\lambda = \lambda_e$ prevent us from 
seeing how this transition happens. Nevertheless, the main 
conclusion from this figure is that the scalar 1-loop correction 
reflects the behaviour of the full \ac{QED} 1-loop correction 
to within a factor of about 2 over a wide range of flux tube widths 
for isolated flux tubes.

Besides using a finite field profile, the fermion problem can also be
circumvented by increasing the electron mass.  The square of the electron mass
sets the scale for the exponential suppression of the large proper
time Wilson loops that contribute to the fermion problem. However, if
we increase the fermion mass, we are reducing the Compton wavelength
of our theory so that the flux tube lattice is no longer dense in
terms of the modified Compton wavelength. It is the Compton wavelength of
the theory that determines what is meant by `dense'. We therefore
cannot avoid the fermion problem for dense lattice models by changing
the electron mass.

Based on the results presented in figure \ref{fig:scalfermiratio}, we
conclude that the coupling between the electron's spin and the
magnetic field do not have a dramatic effect on the vacuum energy for
isolated flux tubes. Therefore, we expect that \ac{ScQED} provides a
good model of the underlying vacuum physics near these flux tubes, at
least at the level our toy model flux tube lattice.

\subsection{Flux Tube Lattice}

The worldline numerics technique computes an effective action density which is then integrated to 
obtain the effective action. This quantity differs from the Lagrangian in that it is 
not determined by local operators, but encodes information about the field everywhere
through the worldline loops.
Like the classical action, the 1-loop term of the effective action per unit length 
is infinite for a flux tube lattice because the field extends infinitely far. 
For this reason, we define the effective action to be the action density integrated 
over the region of a central flux tube $(0<\rho<a/2)$:

\small
\ba
	\frac{\Gamma}{\mathcal{T}L_z} &=& -\pi \int_0^{a/2}\rho B_z(\rho)^2d\rho - \nonumber \\
	& &\frac{1}{2\pi}\int_0^{a/2}\rho_{\rm cm} d\rho_{\rm cm}\int_0^\infty \frac{dT}{T^3}e^{-m^2T} \times
	\nonumber \\
	& & \left\{ \mean{W}_{\rho_{\rm cm}} - 1 +\frac{1}{6}(eB(\rho_{\rm cm})T)^2\right\}.
\ea
\normalsize

\begin{figure}
	\centering
		\includegraphics[width=\linewidth,clip,trim=0.4in 0.1in 0.55in 0.3in]{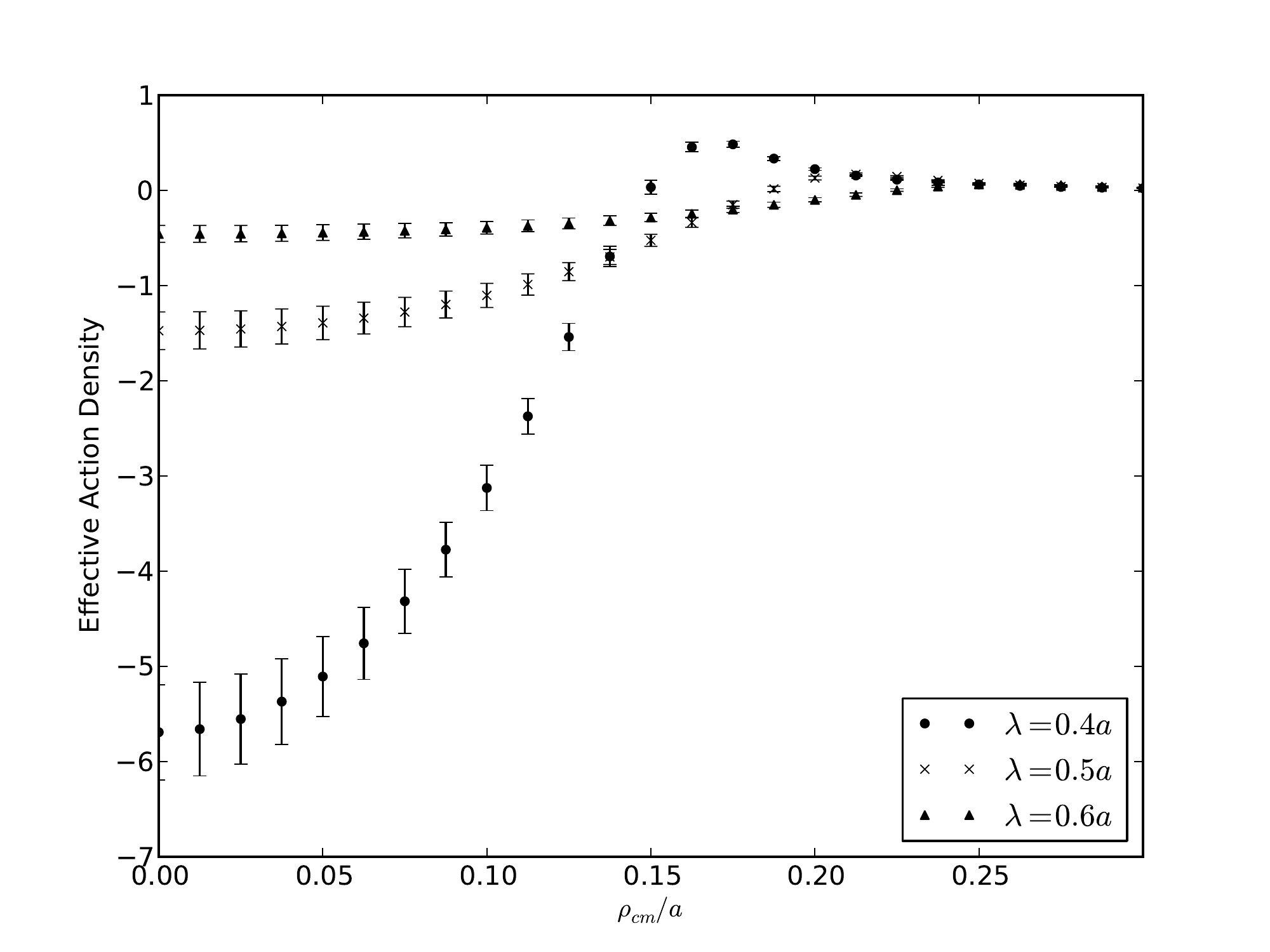}
	\caption[The 1-loop action density for a cylindrical lattice in \acs{ScQED}.]
	{The \ac{ScQED} effective action density for the central 
	flux tube in our cylindrical lattice model for several tube widths, 
	$\lambda$. The average field strength is the critical field, $B_K$. 
	The effective action is positive in between flux tubes due to non-local 
	effects.}
	\label{fig:EAscffpeek}
\end{figure}

The 1-loop term of the effective action density is plotted in figure
\ref{fig:EAscffpeek} for the cylindrical flux tube lattice model. The
most pronounced feature of this density is that there is a negative
contribution from the regions where the field is strong.  This
contribution has the same sign as the classical term. Therefore,
the quantum correction tends to reinforce the classical action. A less
pronounced feature is that there is a positive contribution arising
from the $\rho_{\rm cm} > \lambda/2$ region, in between the lumps of magnetic
field which represent the flux tubes.  In this region, the local
magnetic field is positive, but small.

\begin{figure}
	\centering
		\includegraphics[width=\linewidth,clip,trim=0.5in 0.1in 0.55in 0.3in]{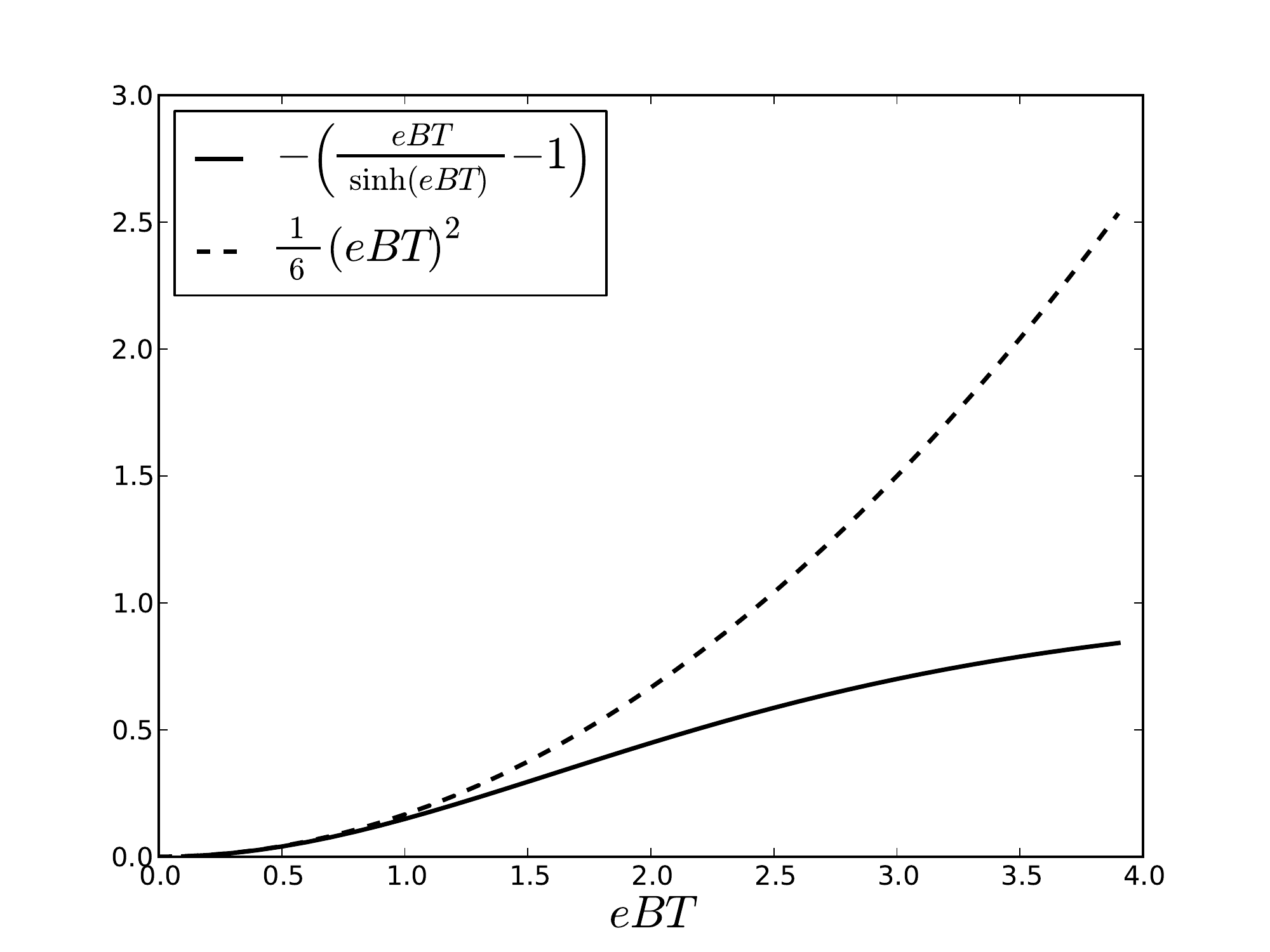}
	\caption[Wilson loop and counterterm contributions for a constant field]
	{The Wilson loop and counterterm contributions to the integrand 
	of the effective action for a constant field in \ac{ScQED}. For constant fields, the 
	effective action is always negative due to the domination of the counterterm 
        over the Wilson loop. For non-homogeneous fields, 
	a positive effective action density signifies 
	that non-local (\ie $T > 0$) effects dominate the counterterm.}
	\label{fig:CounterTermDomination}
\end{figure}

To interpret this feature, we consider the relative contributions between the 
Wilson loops and the counterterm. These terms are shown in figure 
\ref{fig:CounterTermDomination} for the constant field case. 
For all values of proper time, $T$, the counterterms dominate, giving 
an overall negative sign. In order for the action density to be positive, there 
must be a greater contribution from the Wilson loop average than from the 
counterterm, since this term tends to give a positive contribution to the 
action. In our flux tube model, this seems to occur in the regions between the 
flux tubes. In these regions, the local contribution from the counterterm is 
relatively small because the field is small. However, the contribution from the 
Wilson loop average is large because the loop cloud is exploring the nearby 
regions where the field is much larger. The effect is largest where the field 
is small, but becomes large in a nearby region. We therefore interpret the positive 
contributions to the 1-loop correction from these regions as a non-local 
effect. A similar example of such an effect from fields which vary on 
scales of the Compton wavelength has been observed previously using the 
worldline numerics technique~\cite{PhysRevD.84.065035}. 

\begin{figure}
	\centering
		\includegraphics[width=\linewidth,clip,trim=0.15in 0.1in 0.55in 0.3in]{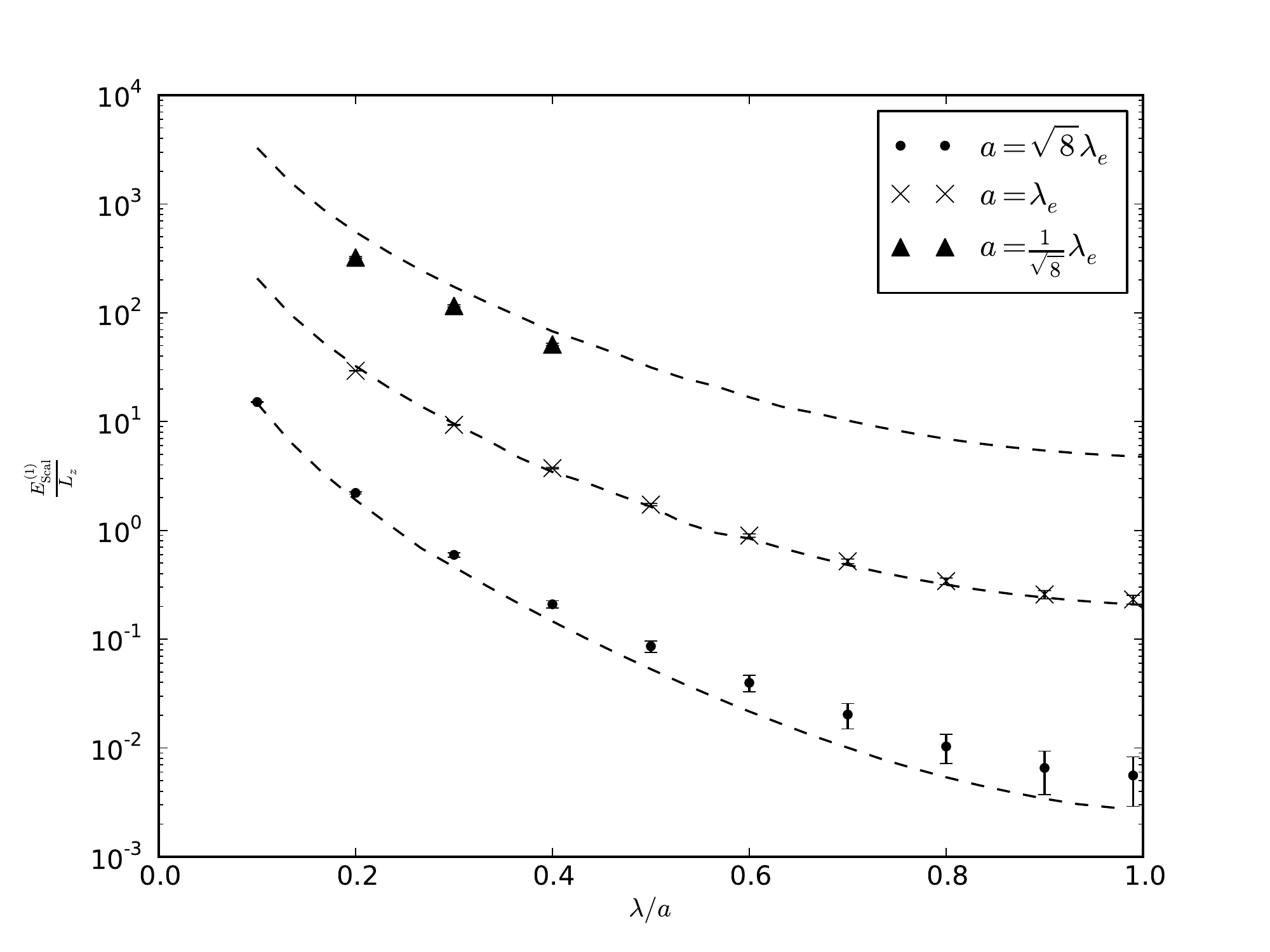}
	\caption[The 1-loop \acs{ScQED} term versus flux tube width]
	{The 1-loop \ac{ScQED} term of the effective action as a function 
	of flux tube width, $\lambda/a$, for several values of the flux tube spacing, $a$.
	The dotted lines are computed from the \ac{LCF} approximation.}
	\label{fig:EAscffvsl}
\end{figure}
\begin{figure}
	\centering
		\includegraphics[width=\linewidth,clip,trim=0.15in 0.1in 0.55in 0.3in]{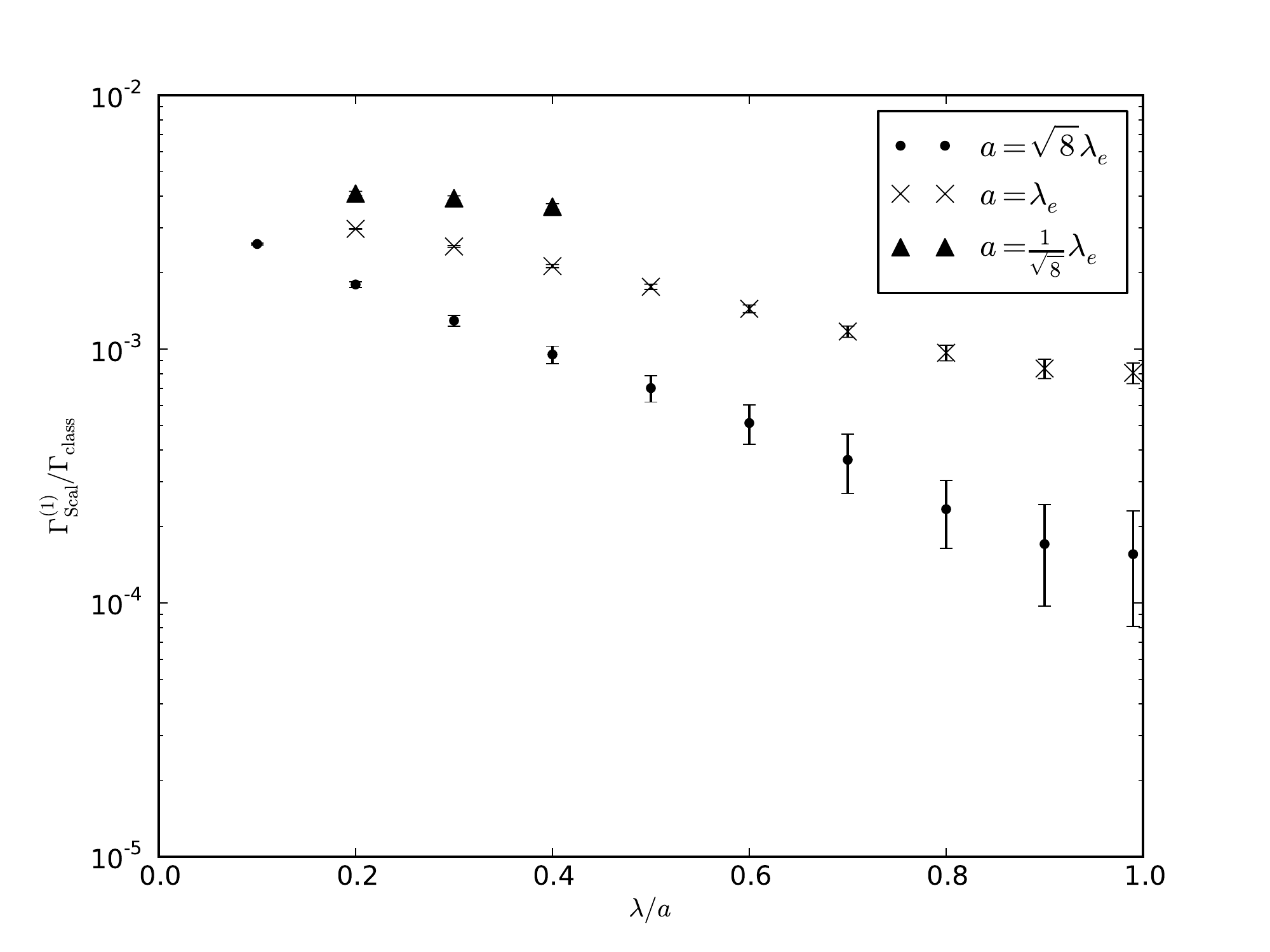}
	\caption[The 1-loop \acs{ScQED} quantum to classical ratio versus flux tube width]
	{The 1-loop \ac{ScQED} term divided by the classical term of the effective action as a function 
	of flux tube width, $\lambda/a$ for several values of the flux tube spacing, $a$.}
	\label{fig:ActRatscff}
\end{figure}

In figure \ref{fig:EAscffvsl}, we plot the magnitude of the 1-loop
\ac{ScQED} term of the effective action as a function of the flux tube
width. As the flux tubes become smaller, there is an amplification of
the 1-loop term, just as there is for the classical action.
Similarly, for more closely spaced flux tubes, $a$ is smaller, and the
1-loop term increases in magnitude. The ratio of the 1-loop term to
the classical term is plotted in figure \ref{fig:ActRatscff}. The
quantum contribution is greatest for closely spaced, narrow flux
tubes, but does not appear to become a significant fraction of the
total action.
\begin{figure}
	\centering
		\includegraphics[width=\linewidth,clip,trim=0.25in 0.1in 0.55in 0.3in]{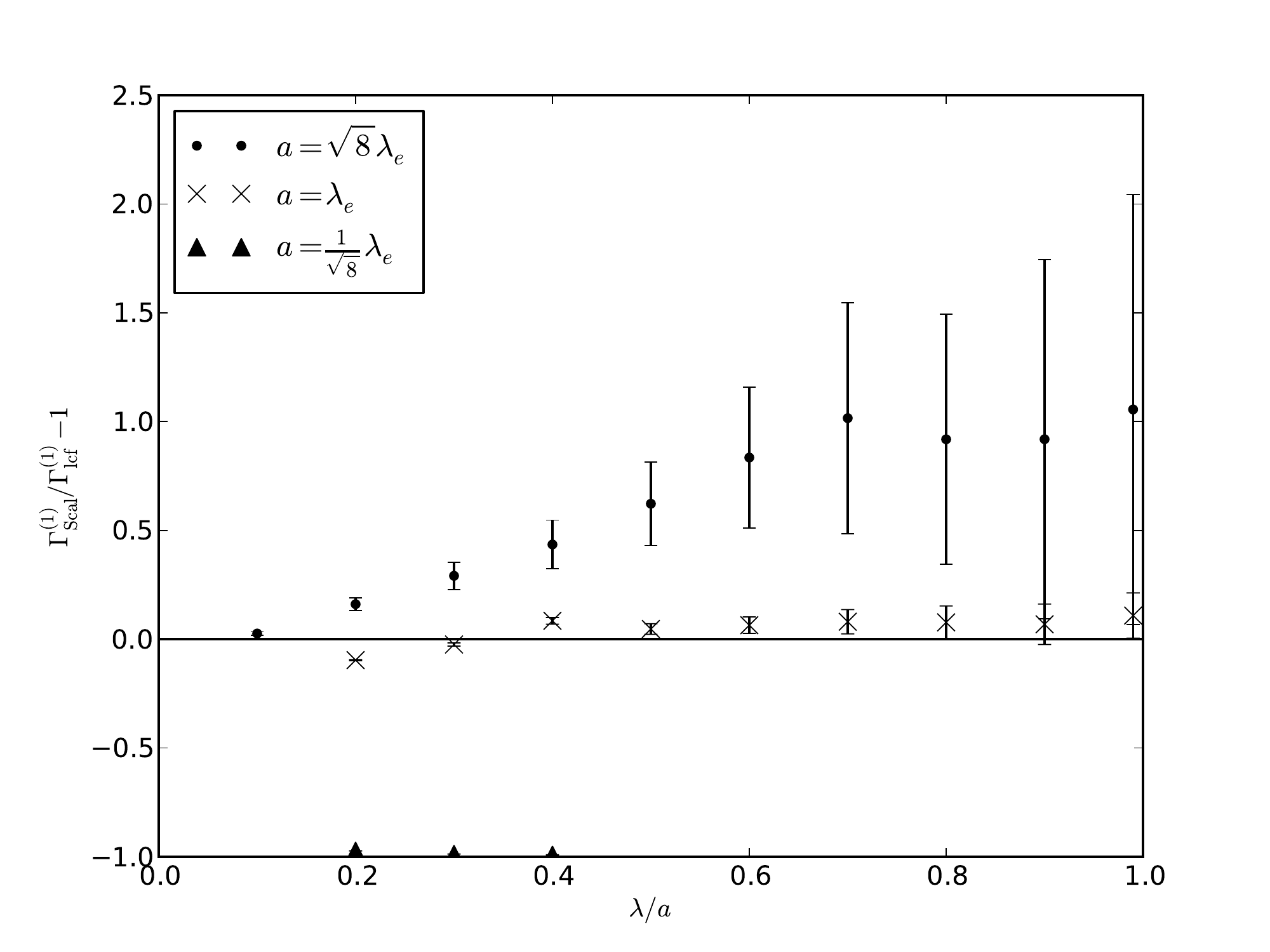}
	\caption[The deviation of the cylindrical flux tube lattice from the \ac{LCF}
	approximation]
	{The residuals between the worldline numerics results and the \ac{LCF} approximation 
	for the cylindrical flux tube lattice. The level of agreement observed is not expected because 
	the field varies rapidly on the Compton wavelength scale. This agreement is believed to 
	be due to an averaging effect of integrating over the electron degrees roughly reproducing 
        the result of a mean-field approximation.}
	\label{fig:EAscffresid}
\end{figure}

We observe that the \ac{LCF} approximation is surprisingly good despite the 
fact that the magnetic field is varying rapidly on the Compton wavelength scale of the electron.
We plot the residuals showing the deviations between the worldline numerics results and the \ac{LCF}
approximation in figure \ref{fig:EAscffresid}. To understand this, recall the discussion 
surrounding figure \ref{fig:CounterTermDomination}. The Wilson loop term is sensitive to the 
average magnetic field through the loop ensemble, $\mean{B}_{\rm e}$. In contrast, the counterterm 
is sensitive to the magnetic field at the center of mass of the loop, $B_{\rm cm}$. 
Since these terms carry opposite signs, we can understand the difference from the 
constant field approximation in terms of a competition between these terms. When 
$B_{\rm cm} < \mean{B}_{\rm e}$, such as when the center of mass is in a local minimum of the 
field, there is a reduction of the energy relative to the locally constant field case, with 
a possibility of the quantum term of the energy density becoming negative. 
However, when $B_{\rm cm} > \mean{B}_{\rm e}$, such as in a local maximum of the field, 
there is an amplification of the energy relative to the constant field case. We can put a 
bound on the difference between the mean field through a loop and the field at the 
center of mass for small loops (\ie small $T$),
\be
	|\mean{B} - B_{\rm cm}| \lesssim |B''(\rho_0)| T
\ee
where $|B''(\rho_0)| \ge B''(\rho)$ for all $\rho$ in the loop. This expression is proved the 
same way as determining the error in numerical integration using the midpoint rectangle 
rule.

If the field varies rapidly about some mean value 
on the Compton wavelength scale, the various contributions 
from local minima and local maxima are averaged out and the mean field approximation 
provided by the \ac{LCF} method becomes appropriate. A similar argument applies in the 
fermion case, where the important quantity is the mean magnetic field along the circumference 
of the loop. This quantity is also well served by a mean-field approximation when integrating 
over rapidly varying fields.

Another interesting feature of figure \ref{fig:EAscffresid} is that the \ac{LCF} 
approximation appears to describe narrower flux tubes better than wider ones, even when the 
spacing between the flux tubes is held constant. This effect is likely a result of the compact 
support given to the flux tube profiles. For narrow tubes, we are guaranteed to have many more 
center of mass points outside the flux tube than inside, giving a smaller energy contribution than 
for isolated flux tubes without compact support where the distinction between inside and outside is 
not as abrupt. This also explains why narrow, closely spaced tubes produce a lower energy than 
is predicted by the \ac{LCF} approximation.

\begin{figure}
	\centering
		\includegraphics[width=\linewidth,clip,trim=0.25in 0.1in 0.55in 0.3in]{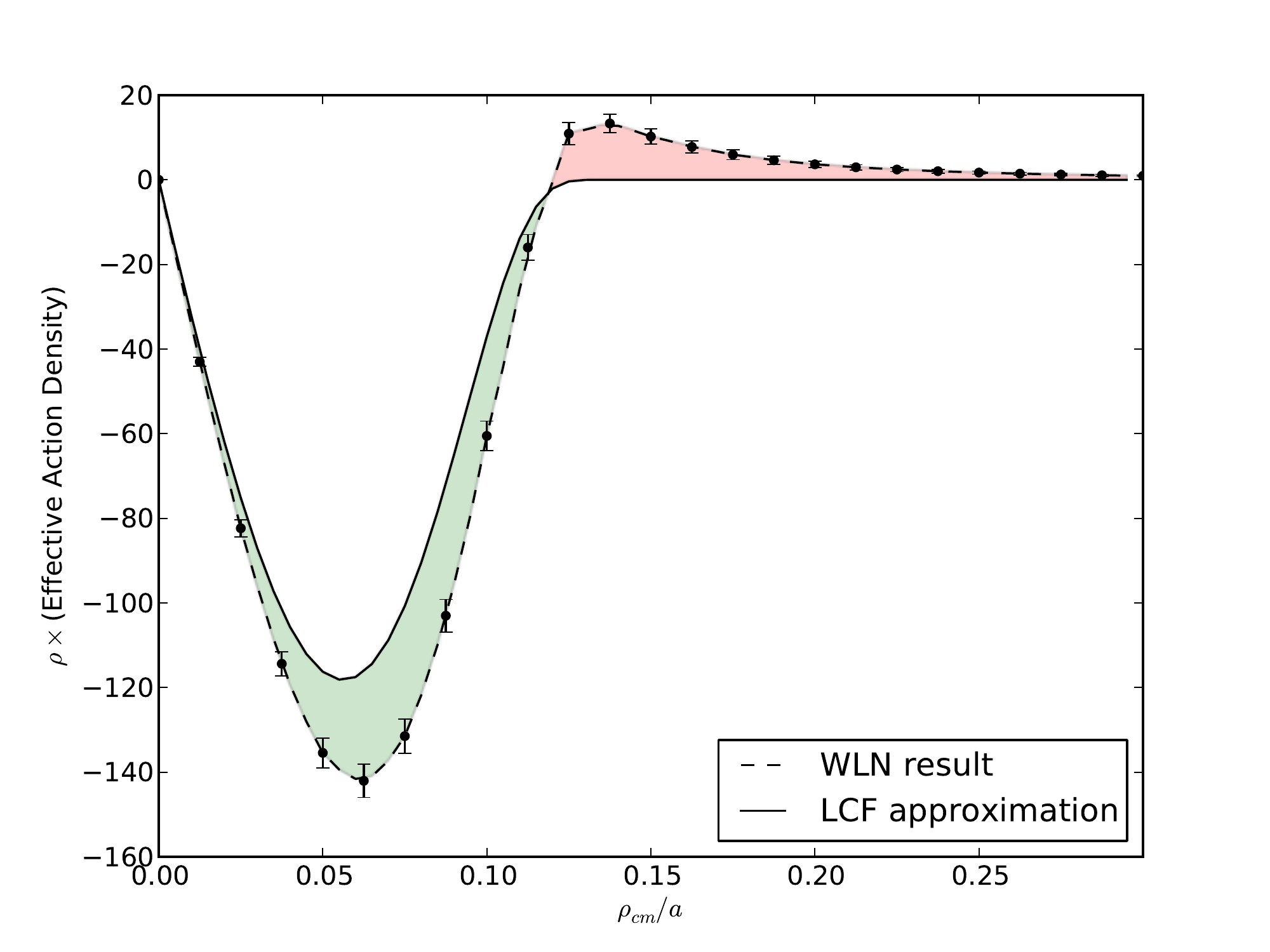}
	\caption[A comparison between the action density in the \ac{LCF} approximation and in worldline numerics]
	{The action density in worldline numerics and the \ac{LCF} approximation, scaled by $\rho$ so 
	an area on the figure is proportional to a volume. The approximation 
	is poor everywhere, however, when there are regions of local minima and local maxima of the field about 
	a mean value, the effective action approximately agrees between these methods. This is due to 
	a partial cancellation between regions where the estimate provided by the approximation is too 
	large (green) and other regions where the estimate is too small (red).}
	\label{fig:EAscffvsconst}
\end{figure}

This argument does not apply to the smooth isolated flux tubes given 
by equation (\ref{eqn:isoB}). 
For these flux tubes, the only region where there is a large discrepancy between $\mean{B}_{\rm e}$ and
$B_{\rm cm}$ is near the center of the flux tube. This 
is a global maximum of the field, and the only maximum of $|B''(\rho)|$. 
There are no regions where the average field in the loop ensemble is 
much stronger than the center of mass magnetic field. So, we expect an amplification of the 
energy near the flux tube relative to the constant field case. 
In the flux tubes with compact support, however, 
there is such a region just outside the flux tube. We can understand the surprisingly close agreement 
of these results to the \ac{LCF} approximation in our model 
in terms of competition between these 
regions of local minima and maxima of the field (see figure \ref{fig:EAscffvsconst}).

Finally, we find that the quantum term remains small compared to the 
classical action for the range of parameters investigated. This is shown 
in figure \ref{fig:ActRatscff} where we plot the ratio of the \acs{ScQED}
term of the action to the classical action. The relative smallness of this correction 
is consistent with the predictions from homogeneous fields and the derivative 
expansion, as well as with previous studies on flux tube 
configurations~\cite{1999PhRvD..60j5019B}.

\subsection{Interaction Energies}

Using this model, we can study the energy associated with
interactions between the flux tubes. Since the flux tubes in our model
exhibit compact support, the interaction energy is entirely due to a
non-local interaction between nearby flux tubes. Thus, it contrasts
with previous research which has investigated the interaction energies
between flux tubes which have overlapping
fields~\cite{Langfeld:2002vy}. In this case, there is a classical
interaction energy ($\propto B_1 B_2$) as well as a quantum correction
($\propto (B_1 + B_2)^4 - B_1^4 - B_2^4$ in the weak-field limit).
Even when these field overlap interactions are not present, there are
also non-local energies in the vicinity of a flux tube due to the
presence of other flux tubes. For example, the energy from nearby flux
tubes can interact with a flux tube through the quantum diffusion of
the magnetic field. Because of this phenomenon, we expect an
interaction energy in the region of the central flux tubes due to the
proximity of neighbouring flux tubes, even though no changes are made
to the field profile or its derivatives in the region of interest.
Since this interaction represents a force due to quantum fluctuations
under the influence of external conditions, it is an example of a
Casimir force. The Casimir force between two infinitely thin flux
tubes in \ac{ScQED} has previously been found to be
attractive~\cite{duru1993}. Our model can shed light specifically on
this interaction, which is not predicted by local approximations such
as the derivative expansion.

Consider a central flux tube with a width $\lambda = 0.5\lambda_e$.
When $\lambda_{\rm min} = \lambda$, the magnetic field outside of the
flux tube, $B_{\rm bg}$, is zero. Then, if the distance between flux
tubes, $a$, is set very large, the energy density will be localized to
the central flux tube and there will be no non-local interaction energy
due to neighbouring flux tubes.  We define the interaction energy,
$E_{\rm int}$, as the difference in energy within a distance $a/2$ of
the central flux tube between a configuration with a given value of
$a$ and a configuration with $a = \infty$. In practice, we use $a =
10,000\lambda_e$ as a suitable stand-in for $a = \infty$:
\be
	\frac{E_{\rm int}}{L_z} = -\frac{\Gamma_{\rm scal}(a)}{L_z \mathcal{T}} 
	+\frac{\Gamma_{\rm scal}(a=1\times 10^4 \lambda_e)}{L_z \mathcal{T}}.
\ee
With this definition, the interaction energy is the energy associated with 
lowering the distance between flux rings from infinity. This the the analogue 
in our model of reducing the lattice spacing of the flux tubes.

One complication of this definition is that there is no clear
distinction between energy density which `belongs' to the central flux
tube and energy density which `belongs' to the neighbouring flux
tubes. We continue to use our convention that the total energy for the
central flux tube is determined by the integral over the non-local
action density in a region within a radius of $a/2$ of the flux
tube. As $a$ is taken smaller and smaller, some energy from nearby
flux tubes is included within this region, but also, some energy
associated with the central flux tube is diffused out of the
region. This ambiguity is unavoidable within this model. We can't
numerically compute the energy over all of space and subtract off
different contributions, because these energies are infinite.

\begin{figure}
	\centering
		\includegraphics[width=\linewidth,clip,trim=0.25in 0.1in 0.55in 0.3in]{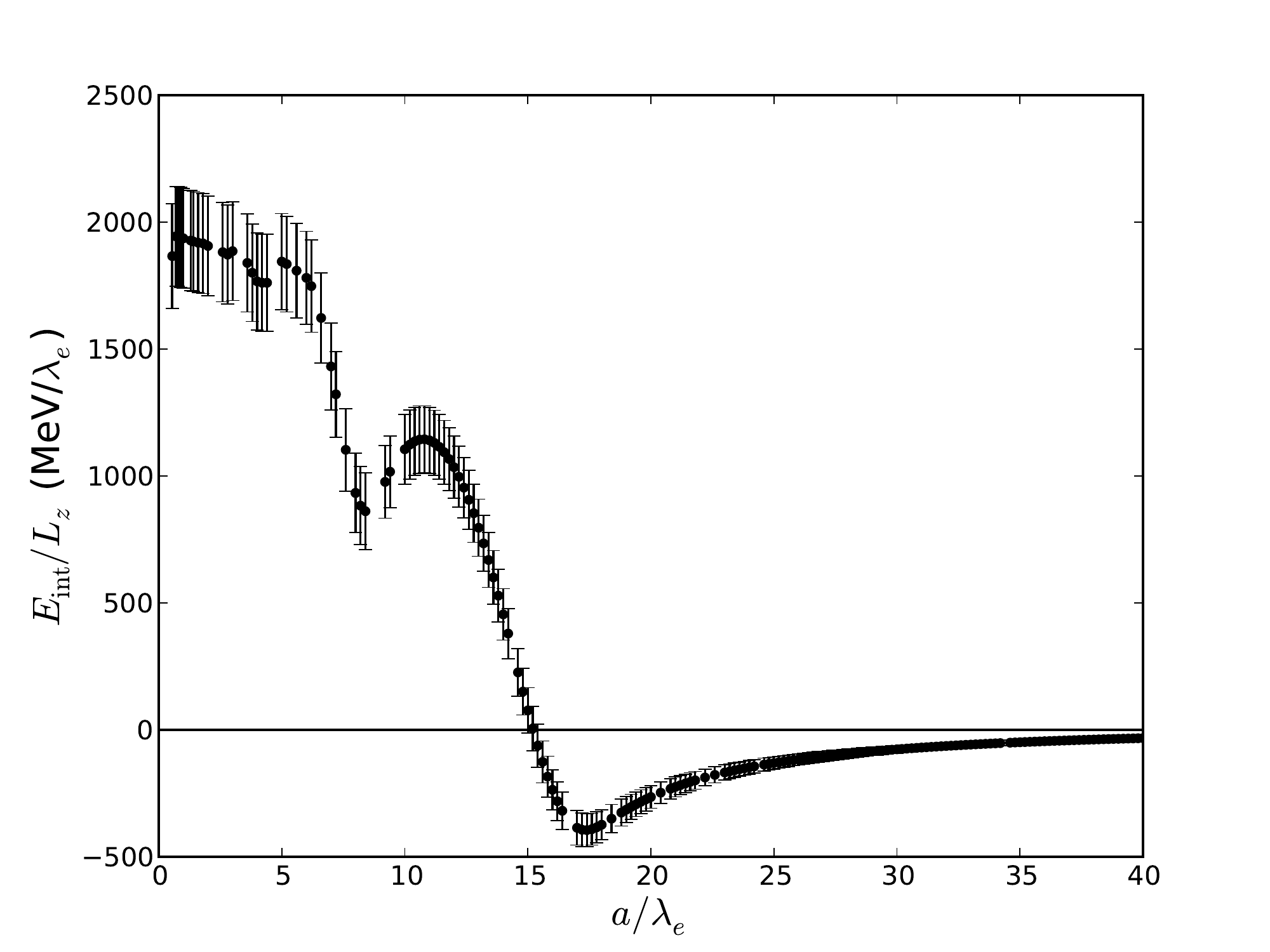}
	\caption[The interaction energy versus the flux tube spacing]
	{The interaction energy per unit length of flux tube as a function 
	of the flux tube spacing, $a$. The energy density of a critical 
	strength magnetic field is 17 GeV/$\lambda_e^3$, so this 
	energy density is small in comparison to the classical magnetic field energies, 
	or the 1-loop corrections. However, the local interactions are constant in $a$.
	At $a<\lambda_e/2$, 
	the bump functions from neighbouring tubes overlap. This approximately corresponds
	to the critical magnetic field which destroys superconductivity.}
	\label{fig:interaction}
\end{figure}

The interaction energy is plotted in figure \ref{fig:interaction}. In
this plot, the error bars are 1-sigma error bars that account for the
correlations in the means computed for each group of worldlines,
\be
	\sigma_{E_{\rm int}} = \sqrt{\sigma_{E_a}^2 + \sigma_{E_{a=10000}}^2 
		- 2\operatorname{Cov}(E_a, E_{a=10000})},
\ee
where $\operatorname{Cov}(a,b)$ is the covariance between random
variables $a$ and $b$.  Recall from figure \ref{fig:EAscffpeek} that
there is a positive contribution to the effective action, and
therefore a negative contribution to the energy from the region
between flux tubes. As we reduce $a$, bringing the flux tubes closer
together, two considerations become important. Firstly, we are
increasing the average field strength meaning there tends to be more
flux through the worldline loops which tends to give a negative
contribution to the interaction energy. Secondly, we are reducing the
spatial volume over which we integrate the energy since we only
integrate $\rho$ from $0$ to $a/2$. This effect makes a positive
contribution to the interaction energy since we include less and less
of the region of negative energy density in our integral.

In figure \ref{fig:interaction}, there appears to be a landscape with both 
positive and negative interaction energies at different values of $a$. 
These appear to be consistent with the interplay between positive and 
negative contributions described in the previous paragraph. This is consistent with 
the usual expectation of attractive Casimir forces~\cite{duru1993}. The dominant contribution 
for the positive energy values is caused by less of the negative energy region contributing 
as the domain assigned to the flux tube is reduced. However, the point at $a/\lambda_e = 2$ is 
negative ($-0.3\pm0.1$MeV/$\lambda_e$) indicating that an attractive interaction from 
nearby flux tubes is dominant. 

At $a/\lambda_e = 0.5$, the flux tubes are positioned right next to
one another, and the negative contribution from the increase in the
mean field appears to be slightly larger than the positive
contribution from the loss of the region of negative energy density
from the integral. Beyond this, the flux tubes would overlap each
other, which approximately corresponds to the critical background
field which destroys superconductivity.  Based on the above
explanation, it appears that the non-local interaction energy between
magnetic fields has a strong dependence on the specific profile of the
classical magnetic field that was used.  This makes it difficult to
predict if it will result in attractive or repulsive forces in a more
realistic model of a flux tube lattice.

The energy density of a critical strength magnetic field is about
$17{\rm GeV}/\lambda_e^3$, so the energy associated with this
interaction is relatively small.  However, there are no other
interactions which affect the energy of moving the flux tubes closer
together when they are separated by many coherence lengths. Here, the
characteristic distance associated with the interaction, $\lambda_e$,
is considerably larger than coherence length or London penetration
depth, so the interactions between flux tubes through the order field
are heavily suppressed.

\section{Conclusion}
\label{sec:conclusion}

In this paper, we have used the worldline numerics numerical technique 
to compute the effective action of \ac{QED} in
non-homogeneous, cylindrically symmetric magnetic fields. The method
uses a Monte Carlo generated ensemble of worldline loops to
approximate a path integral in the worldline formalism. These
worldline loops are generated using a simple algorithm and encode the
information about the magnetic field by computing the flux through the
loop and the action acquired from transporting a magnetic moment
around the loop.  This technique preserves Lorentz symmetry exactly
and can preserve gauge symmetry up to any required precision.

Computing the quantum effective action for magnetic flux tube 
configurations is a problem that has generated considerable interest
and has been explored through a variety of approaches~\cite{Gornicki1990271, 
1998MPLA...13..379S, 0264-9381-12-5-013, PhysRevD.51.810, 1999PhRvD..60j5019B,
PhysRevD.62.085024, 2001PhRvD..64j5011P, 2003PhRvD..68f5026B, 
2005NuPhB.707..233G, 2006JPhA...39.6799W, Weigel:2010pf, Langfeld:2002vy}
(see section \ref{sec:EAisoflux}). 
Partly, this 
is because it is a relatively simple problem for analyzing non-homogeneous 
generalizations of the Heisenberg-Euler action and for exploring 
limitations of techniques such as the derivative expansion. But, this 
is also a physically important problem because tubes of magnetic flux 
are very important for the quantum mechanics of electrons due 
to the Aharonov-Bohm effect, and they appear 
in a variety of interesting physical scenarios such as stellar astrophysics, 
cosmic strings, in superconductor vortices, and quark confinement
\cite{2003PrPNP..51....1G}.

In the present context, we are concerned with the role that magnetic 
flux tubes play in the superconducting nuclear material of compact stars.
In this scenario, the \ac{QED} effects are particularly interesting because 
the magnetic flux tubes, if they exist, are confined to tubes which may be 
only a few percent of the Compton wavelength, $\lambda_{\rm C}$, in radius. Specifically, 
the flux must be confined to within the London penetration depth of the superconducting 
material, which for neutron stars has been estimated to be 80 fm $=$
0.032 $\lambda_e$~\cite{PhysRevLett.91.101101}. Moreover, the 
flux tube density is expected to be proportional to the 
average magnetic field. For a background field near the quantum critical 
strength, $B_k$, such as in a neutron star, the distance between flux tubes 
is comparable to a Compton wavelength. This Compton wavelength scale is 
also the scale at which the non-locality of \ac{QED} becomes important and 
at which powerful local techniques like the derivative expansion 
are no longer appropriate for computing the effective action.

The free energy associated with these flux tubes is a factor in determining 
whether the nuclear material of a neutron star is a type-I or type-II 
superconductor. The free energy of a flux tube is determined by 
looking at the energies associated with the magnetic field, with the 
creation of a non-superconducting region in the superconductor, and 
with interactions between the flux tubes. Flux tubes can only form 
if it is energetically favourable to do so compared to expelling the 
field due to the Meissner effect.  For a lattice of 
flux tubes, there is also an energy contribution from the 
presence of neighbouring flux tubes because of the non-local 
nature of quantum field theory.

The energy of two flux tubes has been previously
computed using worldline numerics methods and for flux tubes with aligned fields, the 
energy is larger than twice the energy of a single flux tube when the flux tubes 
are closely spaced~\cite{Langfeld:2002vy}. This result implies that there is a 
repulsive interaction between the flux tubes due to \ac{QED} effects,
strengthening the likelihood of the type-II scenario in neutron stars. This interaction 
energy increases as the flux tubes are placed closer together, and can have a similar 
magnitude as the \ac{QED} correction to the energy when the flux tubes are closely spaced.

We have developed a cylindrically symmetric magnetic field 
model which reproduces some of the 
features of a flux tube lattice: for a given central flux tube, there are 
nearby regions of large magnetic field that interact non-locally, and 
the large flux tube size limit goes to a large uniform magnetic field
instead of to zero field. We have investigated the 1-loop effects from 
\ac{ScQED} in this model using the worldline numerics technique for various 
combinations of flux tube size, $\lambda$, and flux tube spacing, $a$.

In contrast to isolated flux tubes, we find that there are some regions where the 
worldline numerics results are greater than the \ac{LCF} approximation and other 
regions where they are less than the \ac{LCF} approximation. This can be understood 
by thinking of the difference from the \ac{LCF} approximation as a competition 
between the local counterterm and the Wilson loop averages. 
For magnetic fields that vary on the Compton wavelength scale 
about some mean field strength, the \ac{LCF} approximation 
provides a poor approximation of the energy density, but may provide a 
good approximation to the total energy density of the field due to it 
being a good mean field theory approximation to the energy density. 
The appropriateness of the \ac{LCF} approximation in this case
can be understood as an approximate balance between regions where the 
field is a local maximum and the magnitude of the quantum correction to the 
action density is larger than in the constant field case, and regions where 
the field is a local minimum and the quantum corrections predict a smaller 
action density than the constant field case. This 
washing out of the field structure due to non-local effects has also 
been observed in worldline numerics studies of the vacuum polarization 
tensor~\cite{PhysRevD.84.065035}. 

There is a force between nearby magnetic flux tubes due 
to the quantum diffusion of the energy density. This interaction is 
non-local and is not predicted by the local derivative expansion.
It is an example of a Casimir force (\ie a force resulting from 
quantum vacuum fluctuations) and it is computed in a very similar 
way as the Casimir force between conducting bodies in the worldline numerics 
technique~\cite{Gies:2003cv}.
The size of the energy densities involved in this force are small even 
compared to the 1-loop corrections to the energy densities, which are
in turn small compared to the classical magnetic energy density. 

Although this interaction energy is small, 
the interactions between flux tubes in a neutron star due to the 
order field of the superconductor are suppressed because the distance between the 
tubes is considerably larger than the coherence length and 
London penetration depth. Therefore, this force is possibly important 
for the behaviour of flux tubes in neutron star crusts and interiors.
For example, in our lattice model, this force could contribute 
to a bunching of the worldlines, producing regions where flux tubes 
are separated by $\sim2\lambda_e$ and other regions which have 
no flux tubes. Consequently, this force may have important implications 
for neutron star physics. However, investigating these implications is 
outside the scope of this paper.

The nature of this interaction energy is expected to depend on the
model of the magnetic field profile for the reasons discussed in
section \ref{sec:periodic_results}. It is therefore reasonable that
forces of either sign, attractive or repulsive, may be possible
depending on the particular landscape formed by the magnetic field and
the particular definition used of the interaction energy. In a
superconductor, the profiles of the magnetic flux tubes are determined
by the minimization of the free energy for the interacting system
formed by the magnetic and order parameter fields. Therefore,
investigating this phenomenon using more realistic models (\ie two-dimensional
triangular lattices with field profiles 
motivated by the physics of superconductors) is an
interesting direction for future research. In particular, it would be
interesting to determine if certain conditions allowed for a non-local
interaction between magnetic flux tubes to be experimentally
observable despite the small forces involved.

These conclusions are directly applicable to \ac{ScQED}.  However, we
have also investigated the relationship between spinor and scalar
\ac{QED} for isolated flux tubes where the worldline numerics technique can be
applied to both cases. We find that both theories have the same
qualitative behaviour, and agree within a factor of order unity
quantitatively. The arguments and explanations given for the
\ac{ScQED} results have strong parallels in the spinor \ac{QED}
case. The spinor case can also be understood in terms of a competition
between the Wilson loop averages and the local counterterm. We
therefore speculate that the results from this paper will hold in
the spinor case, at least qualitatively. However, addressing the
fermion problem so that the spinor case can be studied explicitly for
flux tube lattices would be valuable progress in this area of
research.

This work was supported by the Natural Sciences and Engineering Research
Council of Canada, the Canadian Foundation for Innovation, the British
Columbia Knowledge Development Fund.  It has made used of the NASA ADS
and arXiv.org.

\bibliographystyle{prsty}
\bibliography{references}

\end{document}